\documentstyle[12pt,epsfig,axodraw]{article}
\advance\textheight by \topskip
\def\pr#1 #2 #3 { {\rm Phys. Rev.}            {\bf #1}   (#2) #3}
\def\prd#1 #2 #3{ {\rm Phys. Rev.}            {\bf D#1}  (#2) #3}
\def\prl#1 #2 #3{ {\rm Phys. Rev. Lett.}      {\bf #1}   (#2) #3}
\def\plb#1 #2 #3{ {\rm Phys. Lett.}           {\bf B#1}  (#2) #3}
\def\npb#1 #2 #3{ {\rm Nucl. Phys.}           {\bf B#1}  (#2) #3}
\def\prp#1 #2 #3{ {\rm Phys. Rep.}            {\bf #1}   (#2) #3}
\def\zpc#1 #2 #3{ {\rm Z. Phys.}              {\bf C#1}  (#2) #3}
\def\epjc#1 #2 #3{ {\rm Eur. Phys. J.}        {\bf C#1}  (#2) #3}
\def\mpl#1 #2 #3{ {\rm Mod. Phys. Lett.}      {\bf A#1}  (#2) #3}
\def\ijmp#1 #2 #3{{\rm Int. J. Mod. Phys.}    {\bf A#1}  (#2) #3}
\def\ptp#1 #2 #3{ {\rm Prog. Theor. Phys.}    {\bf #1}   (#2) #3}
\def\jhep#1 #2 #3{ {\rm J. High Energy Phys.} {\bf #1}   (#2) #3}
\def\jphg#1 #2 #3{ {\rm J. Phys.}             {\bf G#1}  (#2) #3}
\def\cpc#1 #2 #3{ {\rm Comput. Phys. Commun.} {\bf #1}   (#2) #3}

\newcommand{\be}{\begin{equation}}
\newcommand{\ee}{\end{equation}}
\newcommand{\br}{\begin{eqnarray}}
\newcommand{\er}{\end{eqnarray}}
\newcommand{\ba}{\begin{array}}
\newcommand{\ea}{\end{array}}
\newcommand{\bi}{\begin{itemize}}
\newcommand{\ei}{\end{itemize}}
\newcommand{\bn}{\begin{enumerate}}
\newcommand{\en}{\end{enumerate}}
\newcommand{\bc}{\begin{center}}
\newcommand{\ec}{\end{center}}

\newcommand{\Dir}{\kern -6.4pt\Big{/}}
\newcommand{\Dirin}{\kern -10.4pt\Big{/}\kern 4.4pt}
\newcommand{\DDir}{\kern -8.0pt\Big{/}}
\newcommand{\DGir}{\kern -6.0pt\Big{/}}

\def\frac#1#2{ {{#1} \over {#2} }}

\def\beq{\begin{equation}}
\def\beeq{\begin{eqnarray}}
\def\eeq{\end{equation}}
\def\eeeq{\end{eqnarray}}
\def\a0{\bar\alpha_0}

\def\b0{\beta_0}

\def\ee{e^+e^-}

\def\lms{\Lambda^{(n_{\rm f}=4)}_{\overline{\mathrm{MS}}}}

\def\slashchar#1{\setbox0=\hbox{$#1$}           
     \dimen0=\wd0                                 
     \setbox1=\hbox{/} \dimen1=\wd1               
     \ifdim\dimen0>\dimen1                        
        \rlap{\hbox to \dimen0{\hfil/\hfil}}      
        #1                                        
     \else                                        
        \rlap{\hbox to \dimen1{\hfil$#1$\hfil}}   
        /                                         
     \fi}                                         %

\def\be{\begin{equation}}
\def\ee{\end{equation}}
\def\bea{\begin{eqnarray}}
\def\eea{\end{eqnarray}}

\def\slash{/\kern -5pt}

\def\ims #1 {\ensuremath{M2_{[#1]}}}

\def\s22w{s_{2W}^2}

\begin{document}

\begin{flushright}
{SHEP-06-10}\\ 
{\today}
\end{flushright}

\vspace*{2.0truecm}
\begin{center}
{\Large \bf
Weak corrections to four-parton processes}
\\[1.5cm]
{\large S. Moretti, M.R. Nolten and D.A. Ross}\\[0.15 cm]
{\it School of Physics and Astronomy, University of Southampton}\\
{\it Highfield, Southampton SO17 1BJ, UK}\\[0.25cm]
\end{center}
\vspace*{0.5cm}
\begin{abstract}
\noindent
We report on a calculation of the `mixed' strong and (purely) weak
corrections through the order $\alpha_{\mathrm{S}}^2\alpha_{\mathrm{W}}$ 
to parton-parton processes in all possible channels at 
hadron colliders entering the single jet inclusive cross section.
At both Tevatron and LHC, such effects are always negligible (below
permille level) in the total integrated cross section whilst they
become sizable in differential rates. Specifically, if
such corrections are defined with respect to the full leading-order
result of   ${\cal O}(\alpha_{\mathrm{S}}^2
+\alpha_{\mathrm{S}}\alpha_{\mathrm{EW}}+\alpha_{\mathrm{EW}}^2)$,
we find that, at the FNAL accelerator, they can reach the $-5\%$ 
benchmark in the jet transverse energy (at the kinematical limit of the 
machine, rendering their detection quite difficult). At the CERN 
collider, in the same observable, they exceed the $-10\%$ level already 
at 1 TeV and can reach $-40\%$  at 4 TeV, kinematic regions where 
such corrections will be comfortably observable for standard luminosity.
In addition, such corrections are somewhat sensitive to the 
factorisation/renormalisation scale choice. 
\end{abstract}
\vspace*{0.4cm}
\centerline{Keywords:~{Hadron Colliders, Electroweak physics, 
Higher-order calculations.}}

\section{Introduction}
\label{sec:intro}

As the centre-of-mass (CM) 
energy of present and future hadron colliders
increases well beyond the Electro-Weak (EW) scale, of ${\cal O}$(100 GeV),
such as at the Tevatron at FNAL ($\sqrt s_{p\bar p}=1.96$ TeV) and  the Large
Hadron Collider (LHC) at CERN ($\sqrt s_{pp}=14$ TeV),
it is clear that the impact of higher order EW corrections, particularly
of the purely Weak (W) component, will become more and more important
phenomenologically, with respect to similar Quantum Chromo-Dynamics
(QCD) effects \cite{Hollik:2004dz}. The reason is twofold. On the one hand, the 
strong coupling `constant', $\alpha_{\mathrm{S}}$, decreases with increasing 
energy faster than the
EW one, $\alpha_{\mathrm{W}}\equiv \alpha_{\mathrm{EM}}/\sin^2\theta_W$
(with $\alpha_{\mathrm{EM}}$ the Electro-Magnetic (EM) coupling
constant and $\theta_W$ the weak angle). On the other hand,
the purely weak part of higher order EW effects produces leading 
corrections of the type $\alpha_{\rm{W}}\log^2({\mu^2}/M_W^2)$, 
wherein $\mu$ represents
some typical energy scale affecting the hard process 
in a given observable, e.g., the partonic CM energy $\sqrt{\hat{s}}$.
For large enough $\mu$ values,  such W effects may be competitive not
only with Next-to-Next-to-Leading-Order (NNLO) (as $ \alpha_{\rm{W}}\approx 
\alpha_{\rm{S}}^2$) but also with NLO QCD corrections (e.g., for
${\mu}=0.5$ TeV, $\log^2({\mu^2}/M_W^2)\approx10$). (Clearly,
in actual calculations, combinatorial effects, colour/flavour 
factors as well as subleading terms will have to appropriately
be accounted for.) These `double logs' are of Sudakov origin and are  
due to a lack of cancellation between virtual and real $W$-emission in
higher order contributions. This is in turn a consequence of the 
violation of the Bloch-Nordsieck theorem in non-Abelian theories
\cite{Ciafaloni:2000df-Ciafaloni:2001vt}. (See
Refs.~\cite{Melles:2001ye,Denner:2001mn} for comprehensive reviews.)

The problem is in principle present also in QCD. In practice, however, 
it has no observable consequences, because of the final averaging of the 
colour degrees of freedom of partons, forced by confinement
into colourless hadrons. This does not occur in the EW case,
where the initial state generally has a non-Abelian charge, such
as in proton-(anti)proton collisions. Besides, these
logarithmic corrections are finite (unlike in
QCD), since $M_W$ provides a physical
cut-off for $W$-emission. Hence, for typical experimental
resolutions, softly and collinearly emitted weak bosons need not be included
in the production cross section and one can restrict oneself to the 
calculation
of weak effects originating from virtual corrections only \cite{Denner-Pozzorini}. 
By doing so, similar
logarithmic effects, $\sim\alpha_{\rm{W}}\log^2({\mu^2}/M_Z^2)$, 
are also generated  by virtual corrections due to $Z$-bosons.
Finally, in some instances, all these purely weak contributions can  be
isolated in a gauge-invariant manner from EM effects which therefore may not
be included in the calculation (as it is the case here). (Notice that
QED corrections, just like QCD ones, are not subject to Sudakov enhancement.) 

Furthermore, physics Beyond the Standard Model (BSM), if due to 
 right-handed weak currents \cite{Taxil:1997kj}, 
contact interactions or compositeness 
\cite{Taxil:1996vf} or new massive gauge bosons $W',Z'$
\cite{Taxil:1998ni,Taxil:1996vs}
may well manifest itself in deviations from the SM in jet
quantities (which eventually emerge as observables of  parton-parton
collisions after 
hadronisation: see \cite{cone}--\cite{Ellis:2001aa} for definitions
and reviews of standard algorithms for hadronisation and jet evolution), 
such as jet transverse energy $E_T$ or di-jet invariant masses $M_{jj}$. 
The lower bounds on typical $W',Z'$ masses from direct searches at 
Tevatron are model dependent but are typically around
$500-600\,\mbox{GeV}$ \cite{15}, 
whilst the LHC can access such massive gauge bosons with
masses up to $4-4.5\,\mbox{TeV}$ \cite{22}.
Whilst typical compositeness limits from Tevatron are very mild, compared
to the EW scale, the LHC can probe  (with planned high luminosity)
jet transverse energies
of about 4 TeV or so, where the possible existence of quark substructures
could be manifest. Moreover, 
a tenfold increase in luminosity, as the one currently been discussed for the 
so-called Super-LHC (SLHC) option, would give access to jets of up to 
$E_T\approx 5$ TeV \cite{Gianotti:2002xx}.

The plan of the paper is as follows. In the next Section we describe the
subprocesses that we have computed in terms of topologies of Feynman diagrams.
Sect.~\ref{sec:calc} illustrates the computational techniques adopted.
Sects.~\ref{sec:Tev} and \ref{sec:LHC} present our results for Tevatron
and LHC, respectively. Sect.~\ref{sec:summa} summarises our work and draws
the main conclusions. Finally, in the appendix, we describe in detail 
the helicity amplitude formalism adopted here, also giving some specific
formulae for (some of) the topologies discussed in Sect.~\ref{sec:procs}.  

\section{The subprocesses}
\label{sec:procs}

In view of what explained in the previous section,  
it becomes of crucial importance to assess
the quantitative relevance of one-loop weak effects
entering via\footnote{Hereafter, the notation $\alpha_{\mathrm{W}}$ is
used to signify purely weak effects only whilst 
$\alpha_{\mathrm{EW}}$ exemplifies the fact that
both weak and EM effects are included at the given order.}
 ${\cal O}(\alpha_{\mathrm{S}}^2\alpha_{\mathrm{W}})$
the fifteen possible $2\to2$ partonic
subprocesses responsible for jet production in hadronic 
collisions\footnote{Note that in our treatment we 
identify the jets with the partons
from which they originate.},
namely\footnote{To clarify our notation: e.g., in the case of process (7),
``$qQ \to qQ$ (same generation)'' refers to, e.g., $ud\to ud$
whereas ``$qQ \to qQ$ (different generation)'' to, e.g., $ub\to ub$.}:
\begin{eqnarray}
g g &\to& q \bar q\\
q \bar q &\to& g g\\
q g &\to& q g\\
\bar q g &\to& \bar q g\\
q q &\to& q q\\
\bar q \bar q &\to& \bar q \bar q\\
q Q &\to& q Q ~({\rm{same~generation}})\\
\bar q \bar Q &\to& \bar q \bar Q ~({\rm{same~generation}})\\
q Q &\to& q Q ~({\rm{different~generation}})\\
\bar q \bar Q &\to& \bar q \bar Q ~({\rm{different~generation}})\\
q \bar q &\to& q \bar q \\
q \bar q &\to& Q \bar Q ~({\rm{same~generation}})\\
q \bar q &\to& Q \bar Q ~({\rm{different~generation}})\\
q \bar Q &\to& q \bar Q ~({\rm{same~generation}})\\
q \bar Q &\to& q \bar Q ~({\rm{different~generation}}),
\end{eqnarray}
with $q$ and $Q$ referring to quarks of different flavours,
 limited to $u$-, $d$-, $s$-, $c$- 
and $b$-type, all taken as massless. 
Whilst the first four
processes (with external gluons) were already computed 
in Ref.~\cite{Ellis:2001ba}, 
the eleven four-quark processes were 
tackled in Refs.~\cite{Moretti:2005aa,Moretti:2005ut}
(see also \cite{LesHouches2005-SM})\footnote{Note that
$gg\to gg$ does not appear through
${\cal O}(\alpha_{\mathrm{S}}^2\alpha_{\mathrm{W}})$ nor do
$qq'\to QQ'$, $\bar q\bar q'\to \bar Q\bar Q'$ and  $q\bar q'\to Q\bar Q'$,
if $q'\neq q$ and $Q'\neq Q$.}.
Furthermore, these four-quark processes 
can have Infra-Red  (IR) divergences, both soft and collinear,
so that gluon bremsstrahlung effects ought to be evaluated to obtain
a finite cross section at the considered order. In addition,
for completeness, we have  also included 
the non-divergent $2\to3$ subprocesses 
\begin{eqnarray}
q g &\to& q q \bar q  \\
\bar q g &\to& \bar q \bar q q\\
q g &\to& q Q \bar Q ~({\rm{same~generation}})\\
\bar q g &\to& \bar q \bar Q Q~({\rm{same~generation}}).
\end{eqnarray}
(See Refs.~\cite{Baur:1989qt,bgs} for
tree-level $\alpha_{\mathrm{S}}\alpha_{\mathrm{EW}}$ 
and $\alpha_{\mathrm{EW}}^2$ interference
effects.) As intimated
in the Introduction, we will instead ignore altogether the contributions
of tree-level $\alpha_{\mathrm{S}}^2\alpha_{\mathrm{W}}$ terms
involving the radiation of $W$- and $Z$-bosons. This presumes that
the adopted jet definition efficiently vetoes against those
gauge bosons decaying inside the jets and the detector coverage
minimises the loss of all their decays.
We will return to this matter later on in the paper.

Before proceeding to describe the calculation we have performed, we
list the `topologies' involved (i.e.,
Feynman graphs without reference to specific couplings or internal masses) and briefly discuss their
most salient phenomenological aspects.

\begin{itemize}
\item{Processes 1 ($gg\rightarrow q\bar{q}$) and 2 ($q\bar{q}\rightarrow gg$) (Figs.~\ref{fig:proc1and2a},~\ref{fig:proc1and2b}):\\
The two processes with a gluon pair in the initial or final state do not exist through ${\cal O}(\alpha_{\rm S}\alpha_{\rm W})$
(as we need at least four quark-gluon vertices) thus
the one-loop weak corrections to $\alpha_{\rm S}^2$ correspond to the leading order weak calculation. The $\alpha_{\rm S}^2\alpha_{\rm W}$ calculation
is IR  finite  and therefore we do not include gluon bremsstrahlung diagrams.}
\item{Processes 3 ($qg\rightarrow qg$) and 4 ($\bar{q}g\rightarrow \bar{q}g$) (Figs.~\ref{fig:proc3and4a}, \ref{fig:proc3and4b}):\\
Comments made for the previous two processes also hold in this case.}
\item{Processes 5 ($qq\rightarrow qq$) and 6 ($\bar{q}\bar{q}\rightarrow \bar{q}\bar{q}$) (Fig.~\ref{fig:proc5and6a}):\\
The full calculation for these processes will have a contribution from order $\alpha_{\rm S}\alpha_{\rm W}$ and requires gluon bremsstrahlung diagrams,
meaning that a subtraction procedure will have to be enforced in order 
to cancel the IR divergences between the real and virtual parts of the calculation
(see later on for further details on the techniques adopted to achieve such a cancellation). Naturally, $W$-exchange is here confined
to self-energy and vertex corrections.}
\item{Processes 7 ($qQ\rightarrow qQ$ (same generation)) and 8 ($\bar{q}\bar{Q}\rightarrow \bar{q}\bar{Q}$ (same generation)) (Fig.~\ref{fig:proc7and8a}):\\
Hereafter, notice that we assume a flavour diagonal Cabibbo-Kobayashi-Maskawa (CKM) mixing matrix for quarks. The approximation is justified by the
unitarity of such a matrix,
the small value of the Cabibbo angle
 and by the fact that contributions from top quarks are negligible in comparison to those from light quarks. Besides,
this approximation considerably simplifies the calculation.}
\item{Processes 9 ($qQ\rightarrow qQ$ (different generation)) and 10 ($\bar{q}\bar{Q}\rightarrow \bar{q}\bar{Q}$ (different generation)) (Fig.~\ref{fig:proc9and10a}):\\
The diagrams that contribute to process 9(10) are a subset of those that contribute to process 7(8). As a consequences of the assumption made
for the CKM matrix, we ignore here the interaction
where a $W$-emission/absorption changes a quark generation index
and as such 
the $W$-mediated graphs  connecting the two fermion lines
that contribute to processes 7 and 8 are ignored in 9 and 10, respectively.
For the same reason, there is no tree-level $\alpha_{\rm S}\alpha_{\rm W}$ contribution here.}
\item{Process 11 ($q\bar{q}\rightarrow q\bar{q}$) (Figs.~\ref{fig:proc11a}, \ref{fig:proc11b}, \ref{fig:proc11c}):\\
This is possibly the most involved process of all, in terms of number of different topologies, though 
$W$-corrections are limited to the case of self-energies and triangles.}
\item{Process 12 ($q\bar{q}\rightarrow Q\bar{Q}$ (same generation)) (Fig.~\ref{fig:proc12a}):\\
The topologies are here a subset of those involved in the previous process, upon replacing appropriately
$Z$- with $W$-exchange between the two fermion lines.}
\item{Process 13 ($q\bar{q}\rightarrow Q\bar{Q}$ (different generation)) (Fig.~\ref{fig:proc13a}):\\
The graphs involved here are a subset of the previous process, as $W$-exchange between the two
fermion lines is not present. In addition, unlike the previous case,
there is no allowed interference  that contributes at order $\alpha_{\rm S}\alpha_{\rm W}$ to this process,
due to the  vanishing of the  colour factors.}
\item{Process 14 ($q\bar{Q}\rightarrow q\bar{Q}$ (same generation)) (Fig.~\ref{fig:proc14a}):\\
$W$-exchange between the two fermion lines here only occurs in annihilation diagrams, unlike
all previous cases, where it takes place in scattering graphs only.}
\item{Process 15 ($q\bar{Q}\rightarrow q\bar{Q}$ (different generation)) (Fig.~\ref{fig:proc15a}):\\
The graphs involved here are a subset of the previous process, as $W$-exchange between the two
fermion lines is not present. 
As in process (13)
there is no allowed interference  that contributes at order $\alpha_{\rm S}\alpha_{\rm W}$ to this process,
as all possibilities have vanishing colour factors.}
\item{Processes 16 ($qg\rightarrow qq\bar{q}$) and 17 ($\bar{q}g\rightarrow \bar{q}\bar{q}q$) (Fig.~\ref{fig:proc16to19a}):\\
These processes are IR finite as there are no possible bremsstrahlung diagrams we can write down at $\alpha_{\rm S}^2\alpha_{\rm W}$ order.}
\item{Processes 18 ($qg\rightarrow qQ\bar{Q}$ (same generation)) and 19 ($\bar{q}g\rightarrow \bar{q}\bar{Q}Q$ (same generation)) 
(Fig.~\ref{fig:proc16to19a}):\\
The same comments as in the previous case are valid here in the context of IR finiteness. Note also that 
there are no processes analogous to processes 18 and 19 with quarks of different generations as there are no 
contributing interferences at the required order.}
\end{itemize}

\section{The calculation}
\label{sec:calc}

Given the large number of 
diagrams involved in the computation,
it is of paramount
importance to perform careful checks. In this respect, we 
should mention that our  expressions 
have been calculated independently
by at least two of us using FORM \cite{Vermaseren:2000nd} and 
that some results have also been reproduced by another
program based on FeynCalc \cite{Kublbeck:1990xc}. (See the Appendix for more details
on the procedure adopted for the tensor reduction and amplitude calculation.) 
We also find reasonable agreement with 
Ref.~\cite{Paolo} in the Sudakov limit, 
i.e., for large invariant masses and transverse momenta of the final state,
provided that the final state particles are fairly central.

As already mentioned, IR divergences  occur when the virtual or 
real (bremsstrahlung) gluon is either soft or collinear with the emitting parton
and these have been dealt with by using the formalism
of Ref.~\cite{Catani:1996vz}, whereby corresponding dipole terms 
are subtracted from the bremsstrahlung contributions in order to render the
 phase space integral free of IR divergences. The
 integration over the gluon phase space of these dipole terms
 was performed analytically in $d$-dimensions, yielding pole terms
which cancelled explicitly against the pole terms of the virtual graphs.
There remains a divergence from the initial state collinear configuration,
which is absorbed into the scale dependence of the PDFs and must be matched
to the scale at which these PDFs are extracted. 
Through the order at which we are
working, it is sufficient to take the LO evolution of the PDFs 
(and thus the one-loop running of $\alpha_{\rm{S}}$).

The self-energy and vertex correction graphs contain Ultra-Violet (UV)
divergences
that have been subtracted here by using the `modified' Dimensional Reduction
(${\overline{\mathrm{DR}}}$) scheme at
the scale $\mu=M_Z$. The use of ${\overline{\mathrm{DR}}}$,
as opposed to the more usual `modified' Minimal Subtraction
(${\overline{\mathrm{MS}}}$) scheme, is forced
upon us by the fact that the $W$- and $Z$-bosons contain axial couplings
which cannot be consistently treated in ordinary dimensional 
regularisation. 
Thus the values taken for the running $\alpha_{\rm S}$ refer to the 
 ${\overline{\mathrm{DR}}}$ scheme whereas the
 EM coupling,
$\alpha_{\mathrm{EM}}$, 
has been taken to be $1/128$ at the above subtraction
point. The one exception to this
renormalisation procedure has been the case of the self-energy insertions
on external fermion lines, which have been subtracted on mass-shell,
so that the external fermion fields create or destroy particle states
with the correct normalisation. 

The top quark entering the loops in reactions with external $b$'s has
been assumed to have mass $m_t=175$ GeV
and width $\Gamma_t=1.55$ GeV. The $Z$-mass used was
$M_Z=91.19$ GeV and was related to the $W$-mass, $M_W$, via the
SM formula $M_W=M_Z\cos\theta_W$, where $\sin^2\theta_W=0.232$.
(Corresponding widths were $\Gamma_Z=2.5$ GeV and $\Gamma_W=2.08$ GeV.)
Also notice that Higgs contributions are not included here, as required
by all quarks being massless.
For the strong coupling constant, $\alpha_{\rm S}$, we have used the 
one- or two-loop expression
with 
$\lms$\footnote{Strictly speaking, we should have amended the values taken for
$\alpha_{\rm S}$ in order to account for the difference between the
 ${\overline{\mathrm{MS}}}$ and ${\overline{\mathrm{DR}}}$
schemes, but this difference is numerically negligible.}
 chosen to match the value required by the 
LO and NLO Parton Distribution Functions
(PDFs) used. The latter were CTEQ6L1 (our default) plus CTEQ6L at LO and CTEQ6M
at NLO \cite{Pumplin:2002vw}, respectively. 

The fully differential cross sections for processes (1)--(19) are obtained numerically in {FORTRAN} 
as follows
\begin{equation}
d\sigma = (2\pi)^4 \delta^4\left(\sum p_i - \sum p_f\right)\frac{1}{2\hat s} 
\left(\prod_f\frac{d^3{\bf p}_f}{(2\pi)^32E_f}\right) |{\cal A}|^2 f_1^p(x_1)f_2^{p(\bar p)}(x_2) dx_1 dx_2,
\end{equation}
wherein $p_{i(f)}=(E_{i(f)},{\bf p}_{i(f)})$ are the initial(final) particle momenta, 
 $ |{\cal A}|^2 $ the amplitude squared averaged(summed) over
initial(final) colours and helicities, $x_i$ the usual Bjorken momentum fractions and 
$f_i^{p(\bar p)}$ the proton(antiproton) PDFs. (For convenience we will refer 
to $f_1^p(x_1)f_2^{p(\bar p)}(x_2) dx_1 dx_2$ -- assuming a summation over 
all possible quark and gluon combinations -- as the parton luminosity.) Finally, 
the integrations over the two- or three-body phase spaces and the $x$'s have been
performed using VEGAS \cite{VEGAS}. (A simple change of kinematic variables yields the 
$E_T$ dependence of the cross section discussed below.)  

\section{Tevatron phenomenology}
\label{sec:Tev}
A detailed discussion for the case of Tevatron, geared to comparisons against
existing data from Run 1 and 2,  can be found in Ref.~\cite{Moretti:2005ut}.
This is mainly a discussion on the impact of ${\cal O}(\alpha_{\mathrm{S}}^2\alpha_{\mathrm{W}})$ one-loop 
virtual corrections on high $E_T$ jet samples, where an excess was initially found
by CDF (but not D0) during Run 1 \cite{Affolder:2001fa}, 
with respect to the NLO QCD predictions 
\cite{Aversa:1988fv}--\cite{Giele:1994gf}. It was eventually
pointed out that a modification of the
gluon PDFs at medium/large Bjorken $x$ \cite{Lai:1996mg} can 
reconcile theory and data, even within current
systematics: see, e.g., \cite{Stump:2003yu}.
In fact, notice that with the most recent PDFs 
(e.g., CTEQ6.1M \cite{Pumplin:2002vw}), the preliminary Run 2
 data also seem to be consistent with NLO QCD,
see \cite{newexcess} for CDF, albeit barely. The question as to whether 
${\cal O}(\alpha_{\mathrm{S}}^2\alpha_{\mathrm{W}})$ effects 
may be required in a future comparison between theory
and data comparisons (or 
in the parameterisation of the PDFs) is still open \cite{Thorne}
and such effects are presently being implemented in the {\tt fastNLO}
 package \cite{fastNLO}.

Here, we complement the results of \cite{Moretti:2005ut} by studying the effects of the 
${\cal O}(\alpha_{\mathrm{S}}^2\alpha_{\mathrm{W}})$ one-loop 
virtual corrections over the full
kinematic range of the final state jets, rather than restrict ourselves
to the specific case of published CDF and D0 data samples.  
Fig.~17
shows the size of the aforementioned corrections
relative to the full LO results (as defined in the caption) for three sets
of CTEQ PDFs \cite{cteq} all taken at the factorisation/renormalisation scale $\mu=\mu_F\equiv\mu_R=E_T/2$, i.e., 
half the jet transverse energy, the standard choice in NLO QCD simulations (see
Refs.~\cite{Aversa:1988fv}--\cite{Giele:1994gf}). Weak effects are basically
independent of the PDFs used and results are negligible
for small jet transverse energy, where the differential cross section
is largest and therefore lead to negligible corrections
in the integrated one. However, the corrections
in the differential cross sections become somewhat relevant
at large jet transverse energy. For example, at the kinematic end $E_T=800-900$ GeV
they can reach the $-5\%$ level. It is however doubtful whether Tevatron can reach
the luminosity necessary to isolate jets at such large jet transverse energies and
even so it is most likely that such higher order effects will be overwhelmed
by systematics. For the current highest reach of Run 2, $E_T\approx 650$ GeV,
the ${\cal O}(\alpha_{\mathrm{S}}^2\alpha_{\mathrm{W}})$ effects studied here amount
to only $-2.5\%$ for $\mu=E_T/2$. If one adopts another 
factorisation/renormalisation scale (see 
Fig.~18), they
can be slightly larger. They increase in excess of $-3\%$ for  $\mu=E_T/4$. For
a fixed scale, e.g., $\mu=M_Z$, they reach  $-4\%$. Notice however that 
for higher jet transverse energies the scale dependence reduces considerably.
Of more relevance numerically is the case when the LO term is defined to be
only of  ${\cal O}(\alpha_{\mathrm{S}}^2)$. As discussed in \cite{Moretti:2005ut},
for certain choices of $\mu$, 
the cumulative effect of EW corrections through ${\cal O}(\alpha_{\mathrm{S}}\alpha_{\mathrm{EW}}
+\alpha_{\mathrm{EW}}^2+\alpha_{\mathrm{S}}^2\alpha_{\mathrm{W}})$ could well reach the
$-10$ to $-15\%$ level for $E_T$ values which will be measurable within Run 2,
with the tree-level and one-loop EW terms being of similar size. Thus, it 
is of paramount importance to establish which terms are included in Monte Carlo (MC) programs used to
interpolate the data. 

Clearly, in the case of Tevatron, the aforementioned logarithmic effects bring very little 
enhancement, as the $E_T$ values that can be probed at such a machine are not much larger
than $M_W$ and $M_Z$. Therefore, despite through 
 ${\cal O}(\alpha_{\mathrm{S}}^2\alpha_{\mathrm{W}})$ there are many 
more diagrams available for channels yielding two jets 
in the final state than via 
${\cal O}(\alpha_{\mathrm{S}}^2)$  or indeed 
${\cal O}(\alpha_{\mathrm{S}}\alpha_{\mathrm{EW}})$ and
${\cal O}(\alpha_{\mathrm{EW}}^2)$\footnote{Notice that subprocesses (5), (6), (9) and (10) are
CKM suppressed in our diagonal approximation through  
${\cal O}(\alpha_{\mathrm{S}}\alpha_{\mathrm{EW}})$.}, the fact that  
the Sudakov regime is not reached here  implies that
the size of the corrections to the LO terms of ${\cal O}(\alpha_{\mathrm{S}}^2)$ is never
much larger than ${\cal O}(\alpha_{\rm W})$ (i.e., with no logarithmic enhancement).
Besides, at FNAL energies, jet production is always dominated by quark-antiquark induced channels, 
so that large corrections typical of processes via
quark-quark/antiquark-antiquark scattering (see later on) do not contribute much to the 
hadronic cross section.

\section{LHC phenomenology}
\label{sec:LHC}

More dramatic results are found at LHC energies. Here, the jet transverse 
energy can be very large so that the corrections now include the Sudhakov enhancement.
 Furthermore,
 there is a considerable increase in quark-quark/antiquark-antiquark luminosity which, owing to  the 
large number of four-fermion Feynman diagrams through one-loop (with respect to tree-level), increases significantly
 the impact of ${\cal O}(\alpha_{\mathrm{S}}^2\alpha_{\mathrm{W}})$  
terms.
We observe that
the additional Feynman diagrams generally
interfere constructively. For example, in our diagonal CKM matrix approximation, 
processes (9)--(10) only count one Feynman diagram
at tree-level (via gluon exchange) whereas at one-loop they are mediated by all graphs
depicted in Fig.~\ref{fig:proc9and10a}. As some of the one-loop amplitudes are Sudakov-enhanced and
some others are not and none of them is gauge-invariant on its own, it is not however possible to quantify 
{\sl a priori} the overall impact of the increased number of graphs even channel by channel.
In general, the larger the number of Sudakov-enhanced amplitudes the larger the corrections are.
There is, however,  a contrasting effect
for the LHC due to the fact that gluon induced processes become dominant 
and these are generally subject to small weak corrections.
This affects 
the overall normalisation but not the Sudakov enhancement. Altogether, 
despite the fact that the correction to the total cross section is still small,
the corrections to the differential one at large $E_T$ are substantial, e.g.,
at $E_T=1$ TeV  one-loop effects 
are already about $-10\%$ and increase further,
up to $-35\%$ at 4 TeV, irrespective of the PDFs  used (over most of the
kinematical range): see 
Fig.~19. 
Also notice that the change of 
factorisation/renormalisation scale has  a  smaller impact
at LHC than at Tevatron, as
confirmed by 
Fig.~20. 

The shape of the corrections in the Figs.~19 and 20 can be understood in terms
of the partonic composition of the complete  ${\cal O}(\alpha_{\mathrm{S}}^2\alpha_{\mathrm{W}})$
sample,  distinguishing between processes initiated by (anti)quarks
only and those which have a gluon component in the initial state, see 
Fig.~21. 
This distinction stems from the fact that in
the case of subprocesses initiated by (anti)quarks only, one also has LO EW effects through 
${\cal O}(\alpha_{\rm{S}}\alpha_{\mathrm{EW}}+\alpha_{\mathrm{EW}}^2)$. 
(In the plot, the label {\tt 
LO SM} identifies the sum of terms of ${\cal O}(\alpha_{\mathrm{S}}^2
+\alpha_{\mathrm{S}}\alpha_{\mathrm{EW}}+
\alpha_{\mathrm{EW}}^2)$.) 
These, however, can only reach a $3\%$ effect 
at  $E_T=1$ TeV, with respect to the ${\cal O}(\alpha_{\rm{S}}^2)$ terms ({\tt LO QCD})
and   a $16\%$ effect 
at  $E_T=4$ TeV.
On the other hand, the  
${\cal O}(\alpha_{\mathrm{S}}^2+\alpha_{\mathrm{S}}^2\alpha_{\mathrm{W}})$ terms
(labelled {\tt NLO weak}) lead to corrections
up to $-40\%$ in  the vicinity of 4 TeV, and even at $E_T=1$ TeV they
amount to $-12\%$.
Thus we observe that the one-loop corrections of order 
$\alpha_{\rm{S}}^2\alpha_{\mathrm{W}}$ dominate the
 tree-level interferences of order
$\alpha_{\rm{S}}\alpha_{\mathrm{EW}}$
owing to the large logarithms. 
 Finally, the plot also shows the contributions from only those 
subprocesses that are not initiated by any gluons (denoted by the label {\tt (qq)}): 
it is clear that at very large $E_T$ (the Sudakov regime)
are these channels that dominate much of the jet phenomenology. 

It is also of interest to understand the different behaviours of the 
 ${\cal O}(\alpha_{\mathrm{S}}^2\alpha_{\mathrm{W}})$ effects in terms of
{each} of the partonic channels involved.  To this end,
we present 
Fig.~22(a)--(c)\footnote{Recall that individual NLO terms need not be positive definite, as
they represent interferences. The total LO + NLO result of course is.}, 
showing 
the contributions to the $E_T$ dependent cross section of subprocesses (1)--(15) 
(plus $gg\to gg$, $qq'\to QQ'$, $\bar q\bar q'\to \bar Q\bar Q'$ and  $q\bar q'\to Q\bar Q'$
at LO only) separately at the LHC. We show the range from $E_T=$100 GeV, which is sufficiently far 
 from poorly modelled threshold effects at $E_T\approx M_W/2$, to 
  $E_T=$800 GeV,
where Sudakov effects start being active in the 
${\cal O}(\alpha_{\mathrm{S}}^2\alpha_{\mathrm{W}})$ corrections.
The main purpose of this plot is to illustrate that, at the LHC, 
unlike the case of the Tevatron, 
the magnitude of the gluon luminosity
 inside the proton  leads to  gluon-initiated processes that can easily compete with the
(anti)quark-initiated ones, even at rather large jet transverse energies. In 
fact, at LO, processes (3), (4) and $gg\to gg$  dominate
over the  total of the (anti)quark-initiated  ones for all $E_T$ values
considered in the 
plot\footnote{The relative importance of 
(anti)quark-initiated processes
 originates from the combination
of the valence quark luminosity, which is always large,
 and a large number of Feynman diagram 
for four-quark processes,
as opposed to a gluon luminosity 
which decreases rapidly with increasing partonic energy 
combined with a small numbers of graphs with external gluons \cite{Ellis:2001ba}.}. 
Tab.~1 
quantifies this, e.g., at $E_T=800$ GeV.
The ${\cal O}(\alpha_{\mathrm{S}}^2\alpha_{\mathrm{W}})$ 
corrections are  particularly large  for subprocesses (7)--(8),
 mainly by virtue
of the large number of diagrams involved at loop level (as intimated earlier), 
with respect
to the full LO case. Process (12)  has a large positive  correction. 
This can be understood from the fact that here the tree-level interference between
strong and weak interactions is forbidden by colour and the leading contribution
comes from a gluon exchanged in the $s$-channel. However, at one-loop level the
cross section is substantially enhanced by the interference between a $t$-channel $W$-exchange
and a $t$-channel gluon exchange, with a further gluon exchanged in the $t$-channel in
order to conserve colour.
Nevertheless this process only makes a small contribution to the complete jet sample.
 Altogether, despite the fact that corrections
to individual channels can be of a few tens of percent, the overall  
${\cal O}(\alpha_{\mathrm{S}}^2\alpha_{\mathrm{W}})$ 
effect is approximately $-6\%$, with respect to the 
${\cal O}(\alpha_{\mathrm{S}}^2\alpha_{\mathrm{S}}\alpha_{\mathrm{EW}}+\alpha_{\mathrm{EW}}^2)$.
Similar patterns for the corrections are typical also at larger jet transverse energies,
with the overall size of the one-loop corrections increasing steadily like 
$\sim\log^2(E_T^2/M_W^2)$ or $\sim\log^2(E_T^2/M_Z^2)$ 
 and the quark-(anti)quark initiated processes gradually
becoming dominant.

\section{Summary and conclusion}
\label{sec:summa}
In summary, whilst at the Tevatron  ${\cal O}(\alpha_{\mathrm{S}}^2\alpha_{\mathrm{W}})$
one-loop virtual corrections to the single jet inclusive cross section may be comparable
to statistical and systematic effects, at the LHC such terms are
important contributions at large jet transverse energies. For the expected highest reach of the 
CERN machine (assuming standard luminosity),
$E_T\approx 4$ TeV, they can be as large as an astounding $-35$ to $-40\%$
(depending on whether these are related to the LO QCD cross section or to the total tree-level
one including interference between strong and EW interactions and the square of the latter).
Therefore, they ought to be included in any comparison of theory with data
in these regimes. However, particular care
should be paid to the treatment of real $W$- and $Z$-production and
decay in the definition of the inclusive jet data sample, as this will determine
whether $W$- and $Z$-bremsstrahlung effects have to be included
in the theoretical predictions through ${\cal O}(\alpha_{\mathrm{S}}^2\alpha_{\mathrm{W}})$,
which might counterbalance the negative effects due to the one-loop
$W$- and $Z$-exchange estimated here \cite{Ciafaloni:2006qu}.
(As these were not included in our calculation, the matter is currently under study
\cite{Joey}.) 

Along the same lines, it should be recalled that 
NNLO EW terms ought to be investigated too, as it is well known from the Sudakov
treatment that they may well be sizable in comparison to the NLO ones (see,
e.g., Ref.~\cite{twoloops}). 

Our results on weak corrections to the single jet inclusive cross section at hadron
colliders are in line with the findings in several other hadronic processes 
and in various approximations (see Refs.~\cite{Beenakker:1993yr}--\cite{Moretti:2006nf}
for an incomplete list limited to the SM). Altogether
they should help to raise the awareness that LHC physics (primarily) is not always
dominated by QCD effects, particularly in extreme kinematic regimes where new physics 
beyond the SM could manifest itself, possibly in observables that are parity violating, 
hence
 sensitive to  genuine  (i.e., of SM origin) EW corrections but not QCD ones.
In our view, progress in evaluating higher order QCD effects should proceed hand-in-hand with that
of assessing similar EW effects.

\section{Appendix}
\subsection{Helicity amplitudes}
It is convenient to consider the virtual corrections in terms of 
helicity amplitudes. This approach has a number of advantages:
\begin{enumerate}
\item Contributions from individual Feynman graphs can be added to the
helicity amplitudes numerically. This allows flexibility for different
 analyses of the various terms entering the virtual corrections as well as
a higher degree of control of the correctness of the results (as more internal
tests can be enforced).
\item The interference with the tree-level amplitudes can also be computed
 numerically. This avoids the cumbersome algebraic expressions that would be 
 obtained if all possible interferences were computed analytically.
\item For applications in which it is possible to polarise the incoming
 beams (e.g., at the Relativistic Heavy Ion Collider (RHIC) \cite{Moretti:2005aa}), 
 the contributions from different helicity combinations can be matched
 with the corresponding polarised PDFs.
\end{enumerate}

The formalism adopted for the case of subprocesses with external gluons,
processes (1)--(4) in Sect.~2, has
already been described in Refs.~\cite{Ellis:2001ba,Maina:2002wz} 
(see also \cite{Maina:2002wz,Maina:2004yc}). For the case of all other channels,
processes (5)--(19) in Sect.~2 (including the bremsstrahlung contributions),
we have found it convenient to adopt a different procedure. In order 
to describe this, let us consider, as an example, the 
amplitude\footnote{Hereafter, $s,t$ and $u$ are
the usual Mandelstam variables at partonic level, 
for which one has $s+t+u=0$ in the case
of massless external particles. For brevity, we have removed here
the `hatted' notation $\hat{s}$, $\hat{t}$ and $\hat{u}$ used elsewhere. }
 ${\cal A}^{(s)}(s,t, \lambda_1,\lambda_2,\lambda_3,\lambda_4)$,
where $\lambda_1, \dots \lambda_4$ are the helicities of the incoming
and outgoing 
partons  as indicated below. 
For process (13) of Sect.~\ref{sec:procs}:
\begin{equation}
q(p_1,\lambda_1) \, + \, \bar{q}(p_2,\lambda_2) \ \to \ 
Q(p_3,\lambda_3) \, + \, \bar{Q}(p_4,\lambda_4),
\label{process}
\end{equation}
where $q$ and $Q$ are quarks of different generations 
(recall that, as mentioned elsewhere, 
to the accuracy to which we are working it is safe to neglect CKM mixing of
flavours in the weak interactions) and $p_i$ ($i=1, ... 4$) their
four-momenta.

The contribution to the amplitude for this process from any Feynman graph
must be of the form
\begin{equation}
{\cal A}^{(s)}(s,t, \lambda_1,\lambda_2,\lambda_3,\lambda_4)
 \ = \ \bar{v}(p_2,\lambda_2) \Gamma_1 u_(p_1,\lambda_1)
  \bar{u}(p_3,\lambda_3) \Gamma_2 v_(p_4,\lambda_4)
 \delta_{\lambda_1,-\lambda_2} \delta_{\lambda_3,-\lambda_4},
\end{equation}
where $\Gamma_1, \ \Gamma_2$ contain strings of $\gamma$-matrices
(as well as propagators and an implied integral over loop
momentum). They are in general tensorial with indices contracted
between $\Gamma_1$ and $\Gamma_2$. 
Since we are interested in energy scales which are much larger than 
the mass of the $b$-quark (and we are not considering $t$-quark production
here), the quarks may be taken to be massless. In this case the strings of
$\gamma$-matrices in $\Gamma_1, \ \Gamma_2$ 
 are vectors for self-energy or vertex correction diagrams 
and two-rank tensors for box diagrams.
They may be reduced so that the
most general forms are
\begin{equation}
\bar{v}(p_2,\lambda_2) \Gamma_1 u_(p_1,\lambda_1) \ = \ 
a_1(s,t,\lambda_1)A_1(s,\lambda_1)  \delta_{\lambda_1,-\lambda_2}
 \, + \, b_1(s,t,\lambda_1)B_1(s\lambda_1)  \delta_{\lambda_1,-\lambda_2},
\end{equation}
with
\begin{equation} A_1(s,\lambda_1) \ = \
 \bar{v}(p_2,\lambda_2) \gamma \cdot w_1 u_(p_1,\lambda_1) \ = \ \sqrt{s} \end{equation}
and
\begin{equation} B_1(s,\lambda_1) \ = \
 \bar{v}(p_2,\lambda_2) \gamma \cdot n u_(p_1,\lambda_1) \ = \
 i\lambda_1\sqrt{s}, \end{equation}
where
\begin{equation} w_1^\mu \ = \ \frac{1}{\sqrt{stu}}\left(up_1^\mu+tp_2^\mu+sp_3^\mu\right)
 \end{equation}
is a unit vector in the scattering plane orthogonal to $p_1$ and $p_2$
and $n^\mu$ is a unit vector normal to the scattering plane.

The coefficients $a_1(s,t,\lambda_1)$ and $b_1(s,t,\lambda_1)$
have the same tensorial structure as $\Gamma_1$ and
can be obtained by taking traces with the appropriate projection
operators, i.e.,
\begin{eqnarray}
a_1(s,t,\lambda_1) & = & \frac{1}{2} {\mathrm{Tr}} \left(
\Gamma_1 \gamma \cdot w_1 \frac{(1-\lambda_1\gamma^5)}{2}\right),
\nonumber \\
b_1(s,t,\lambda_1) & = & \frac{1}{2} {\mathrm{Tr}} \left(
\Gamma_1 \gamma \cdot n \frac{(1-\lambda_1\gamma^5)}{2}\right).
 \end{eqnarray}
The general form for $\bar{u}(p_3,\lambda_3) \Gamma_2 v_(p_4,\lambda_4)$
is obtained similarly
\begin{equation}
\bar{u}(p_3,\lambda_3) \Gamma_2 v_(p_4,\lambda_4) \ = \ 
a_2(s,t,\lambda_3)A_2(s,\lambda_3)  \delta_{\lambda_3,-\lambda_4}
 \, + \, b_2(s,t,\lambda_1)B_2(s\lambda_3)  \delta_{\lambda_3,-\lambda_4},
\end{equation}
where 
\begin{equation} A_2(s,\lambda_3) \ = \
 \bar{u}(p_3,\lambda_3) \gamma \cdot w_2 v_(p_4,\lambda_4) \ = \ -\sqrt{s} \end{equation}
and
\begin{equation} B_2(s,\lambda_3) \ = \
 \bar{u}(p_3,\lambda_3) \gamma \cdot n v_(p_4,\lambda_4) \ = \
 i\lambda_3\sqrt{s}, \end{equation}
with 
\begin{equation} w_2^\mu \ = \ \frac{1}{\sqrt{stu}}\left(tp_3^\mu+up_4^\mu+sp_2^\mu\right).
 \end{equation}

The contribution to the complete helicity amplitude from any graph can therefore
be specified in terms of the coefficients $a_1,b_1,a_2,b_2$. \bigskip

For example, consider the tree-level graph contributing to process 
(13) in eq.~(\ref{process})
due to the exchange of a $Z$-boson,\\

\centerline{\epsfig{file=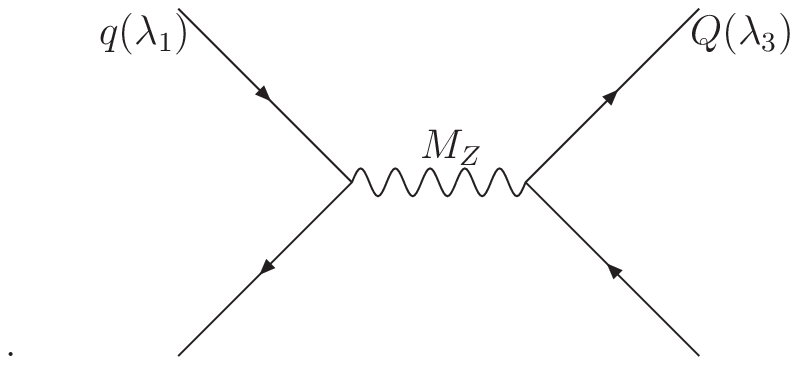, width=6 cm}}
\vspace*{0.25cm}
\noindent
In Feynman gauge, 
we set $\Gamma_1=\gamma^\mu$ and $\Gamma_2=\gamma_\mu$,
and multiply the entire graph by
$$ F \ = \ g^Z_{q,\lambda_1}g^Z_{Q,\lambda_3}
\frac{1}{(s-M_Z^2+i\Gamma_ZM_Z)},$$
where  $g^Z_{q,\lambda_1}  \  g^Z_{Q,\lambda_3}$ are the couplings
of the $Z$ to the quarks $q$ and $Q$, respectively.
We find
\begin{eqnarray}
a_1 & = & F w_1^\mu, \nonumber \\
b_1 & = & F n^\mu, \nonumber \\
a_2 & = & F w_{2\mu}, \nonumber \\
b_2 & = & F n_\mu, \end{eqnarray}
so that the contribution to the helicity matrix element is
\begin{equation}
{\cal A} \ = \ F s \left(w_1^\mu+n^\mu \right)\left(
 -w_{2\mu}+n_\mu \right) \ = \ 
 g^Z_{q,\lambda_1} g^Z_{Q,\lambda_3}\frac{1}{(s-M_Z^2+i\Gamma_ZM_Z)}
 \left[s+2t+s\lambda_1\lambda_3\right] \label{treeamp}. \end{equation}

 The large set of Feynman graphs which contribute to any elementary process
can be classified in terms of a small number of prototype graphs.
The prototype graphs for process (\ref{process}) are listed in the next
section along with their contributions to the helicity matrix element.

For the case of quark-antiquark annihilation in which the quarks $q$ 
and $Q$ are of the same flavour (or at least the same generation), there will
also be contributions to the helicity matrix element from graphs
 involving the exchange of gauge bosons in $t$-channel.
 The contribution from such graphs can be obtained by crossing symmetry
from the graphs with the gauge bosons exchanged in $s$-channel,
\begin{equation}
{\cal A}^{(t)}(s,t,\lambda_1, \lambda_2,\lambda_3 \lambda_4) \ = \ 
 - {\cal A}^{(s)}(t,s,\lambda_1, -\lambda_3, \lambda_2 -\lambda_4).
\end{equation}

Similarly, once we have the helicity matrix element for
quark-antiquark annihilation, the amplitudes for quark-quark
or antiquark-antiquark scattering can be obtained using the
usual crossing relations.

\subsection{Individual topologies}

In this subsection we display the contributions to the total helicity amplitude
of process (13) in eq.~(\ref{process}). 
The graphs are calculated in Feynman gauge in the $\overline{\rm{DR}}$ scheme.
Both IR and UV poles have been subtracted using a
 common subtraction scale, $\mu$. A generic mass $M$ is here attributed 
(for illustration purposes only) to all massive internal gauge bosons. 
Finally, all coupling constants and colour factors have been set to unity
(we indicate the appropriate colour factor in cases where there may be a sign
ambiguity). Notice that the forthcoming contributions have the same overall phase as the
tree-level amplitude given in eq.~(\ref{treeamp}).

\subsubsection{Self-energy graphs}

The contribution to the amplitude from a graph with a self-energy
correction due to a massive gauge boson (mass $M$) on an external leg \\

\centerline{\epsfig{file=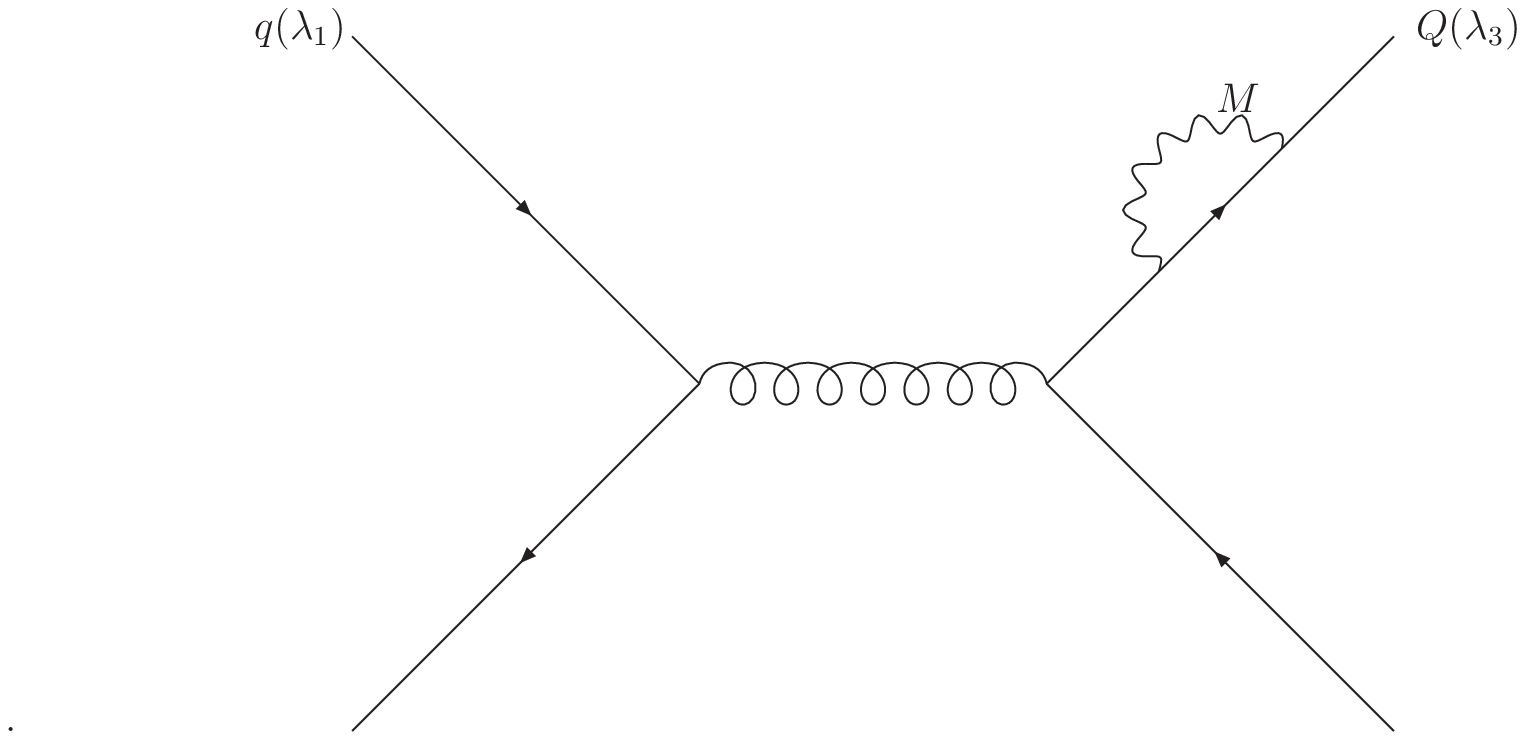, width=7 cm}}

\noindent
is

\begin{equation}\frac{1}{16\pi^2} \left\{ \frac{1}{2}+
\ln\left(\frac{M^2}{\mu^2}\right) \right\}
    \left[1+\frac{2t}{s}+\lambda_1\lambda_3 \right].
\end{equation}
\bigskip

The contribution to the amplitude from a graph with a gluon loop
on the internal gluon line \\

\centerline{\epsfig{file=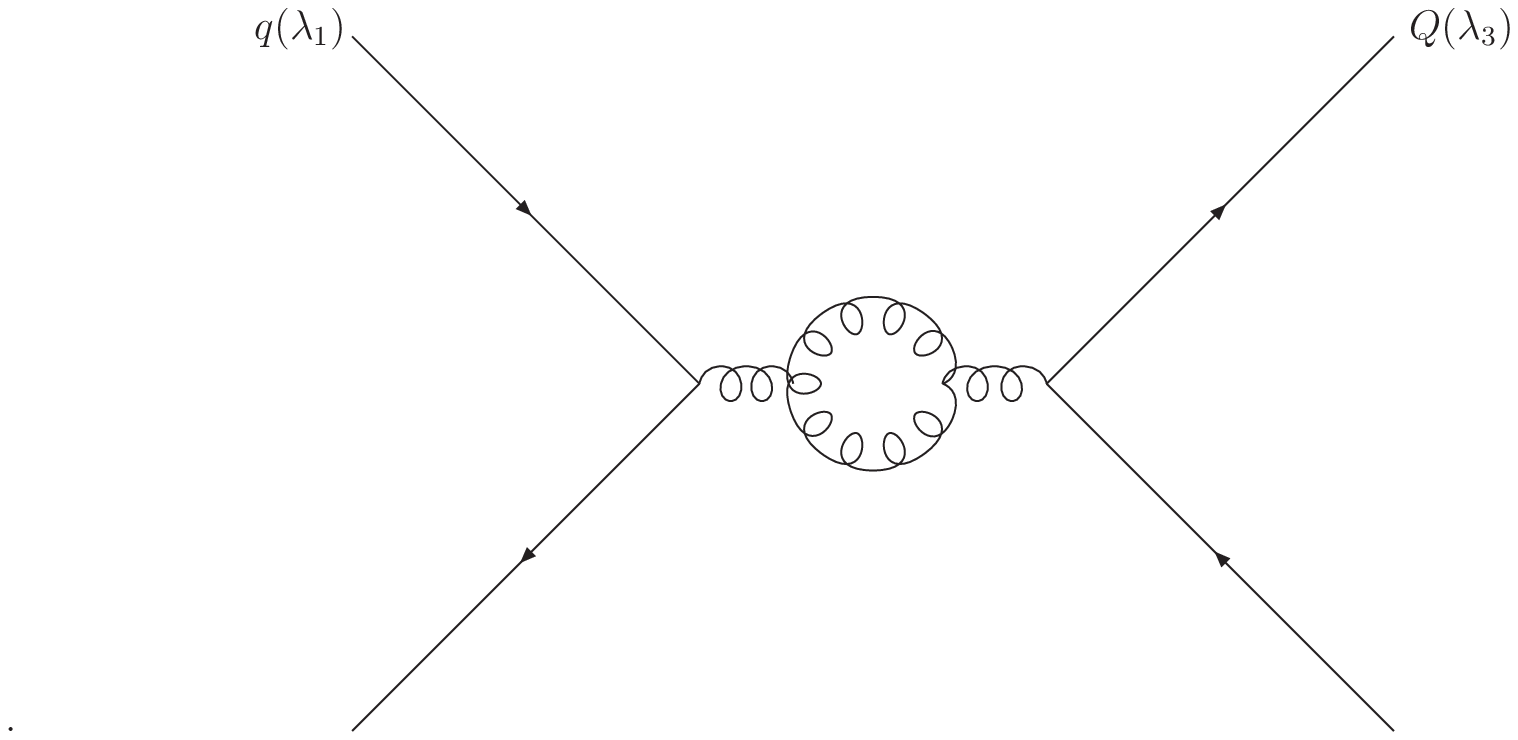, width=7 cm}}

\noindent
is
\begin{equation} -\frac{1}{16\pi^2} \frac{5}{3}
\ln\left(\frac{-s}{\mu^2}\right) 
   \left[1+\frac{2t}{s}+\lambda_1\lambda_3 \right].
\end{equation}
(Here, the relevant colour factor is $C_A$.)
\bigskip

The contribution to the amplitude from a graph with a fermion-loop
self-energy correction to an internal gluon line \\

\centerline{\epsfig{file=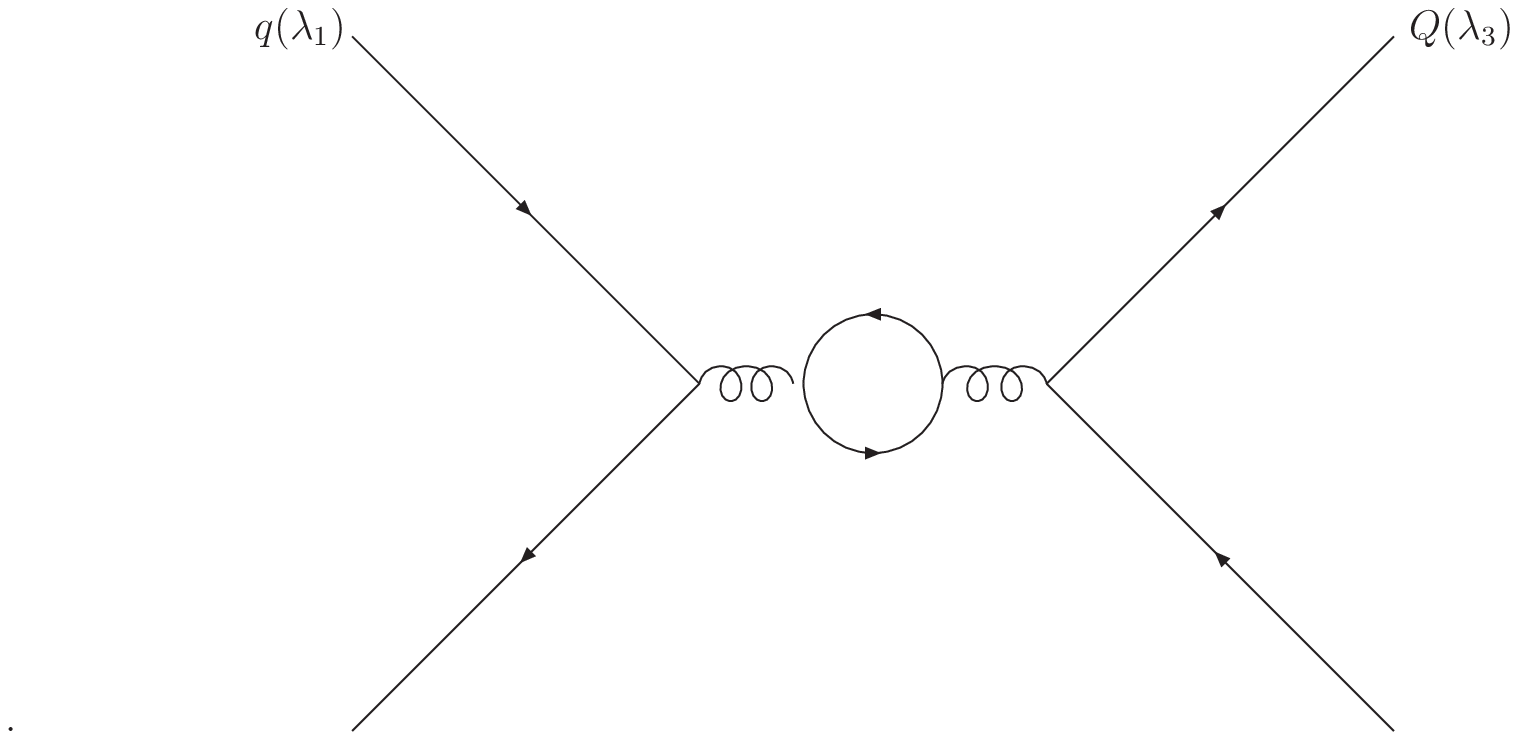, width=7 cm}}

\noindent
is

\begin{equation}\frac{1}{16\pi^2} \frac{4}{3}
\ln\left(\frac{-s}{\mu^2}\right) 
   \left[1+\frac{2t}{s}+\lambda_1\lambda_3 \right].
\end{equation}
(Here, the relevant colour factor is $T_R$.)

\bigskip  \bigskip

\subsubsection{Vertex graphs}

The contribution to the amplitude from a correction to a vertex
due to a massless gauge boson \\

\centerline{\epsfig{file=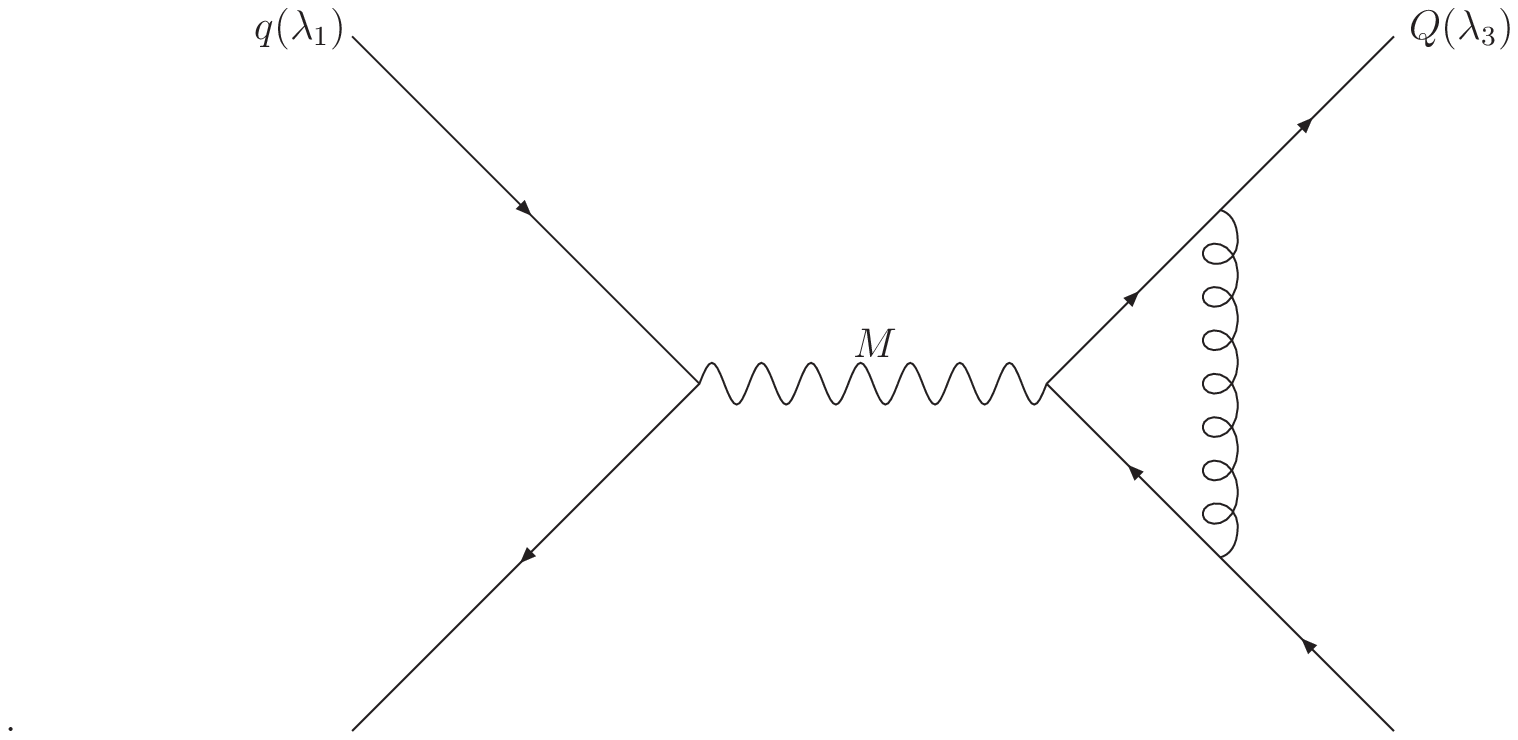, width=7 cm}}

\noindent
is

\begin{equation}
 \frac{1}{16\pi^2}\left\{
 -\ln^2\left(\frac{-s}{\mu^2}\right)+3\ln\left(\frac{-s}{\mu^2}\right)
 +7 \right\}  \frac{s}{(s-M^2)}
 \left[1+\frac{2t}{s}+\lambda_1\lambda_3 \right]. \end{equation}

The contribution to the amplitude from a correction to a gluon vertex
due to a massless gauge boson with triple gauge-boson coupling \\

\centerline{\epsfig{file=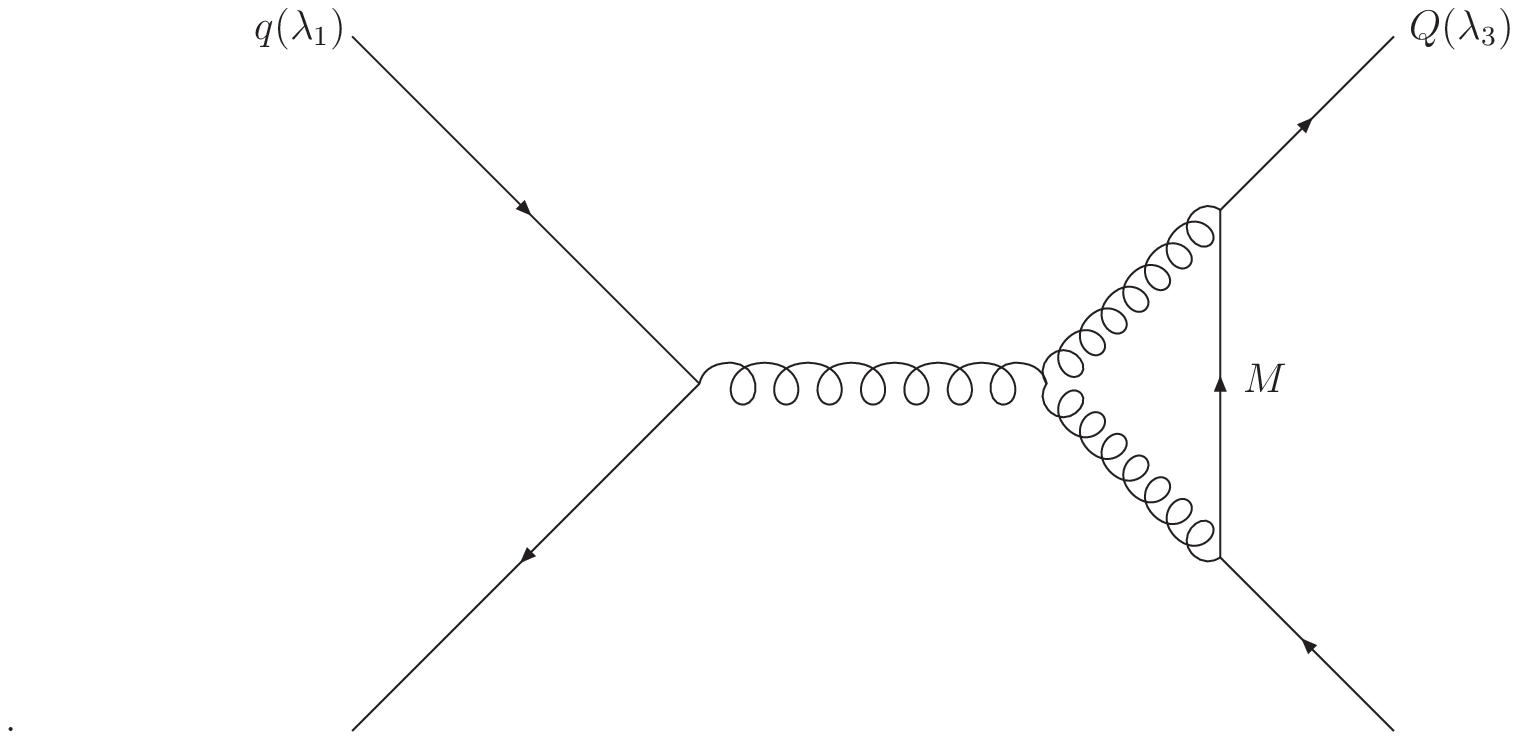, width=7 cm}}

\noindent
is
\begin{equation}
 -\frac{1}{16\pi^2}\left\{ \ln\left(\frac{-s}{\mu^2}\right)
  -1 \right\} 
 \left[1+\frac{2t}{s}+\lambda_1\lambda_3 \right]. \end{equation}
The convention for the triple gauge-boson coupling has been chosen
such that the colour factor associated with this graph is $C_A$.

The contribution to the amplitude from a correction to a QCD vertex
due to a massive gauge boson \\

\centerline{\epsfig{file=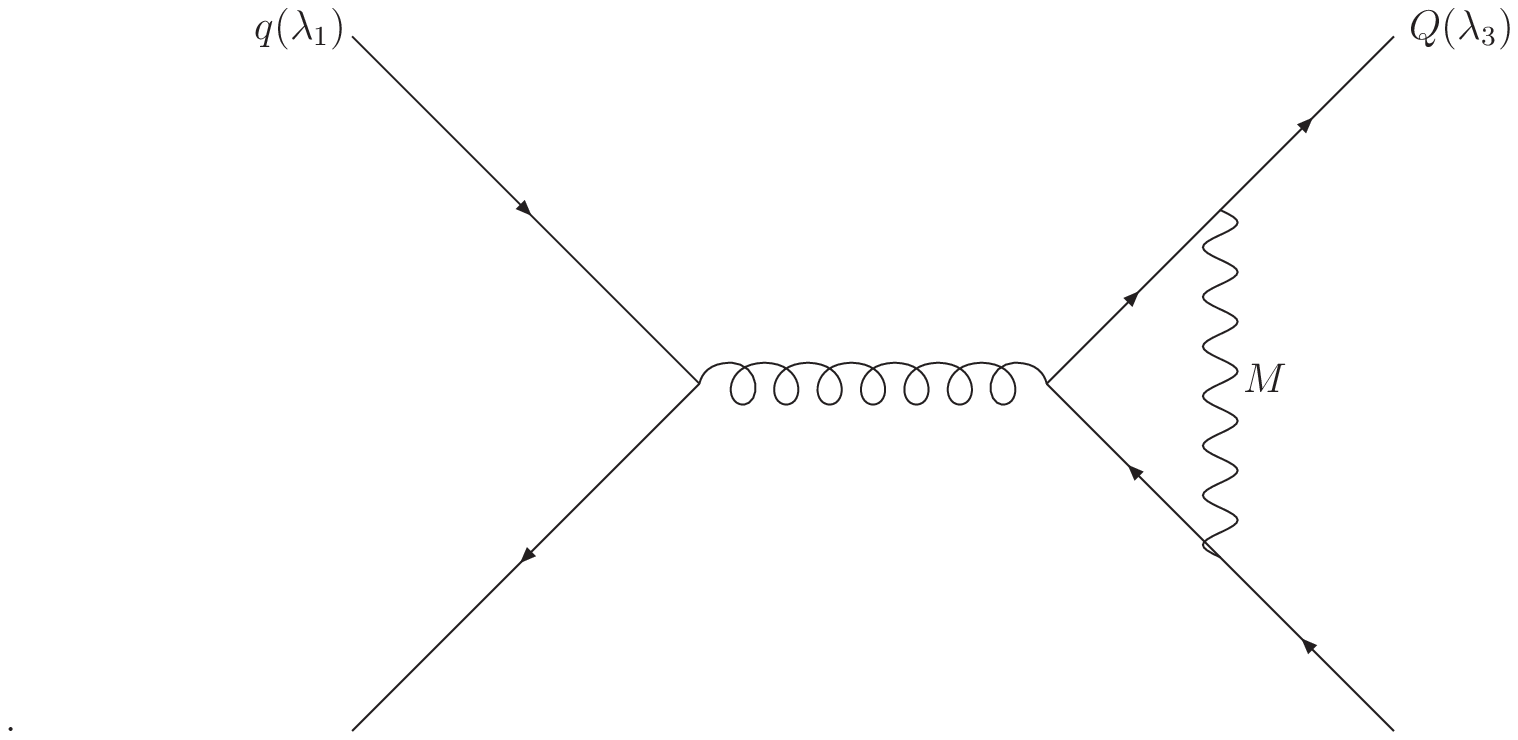, width=7 cm}}

\noindent
is
\begin{eqnarray} \frac{1}{16\pi^2}& \times & \left\{
\left(3+\frac{2M^2}{s}\right)\left[\ln\left(\frac{-s}{M^2}\right)-1\right]
 - \ln\left(\frac{M^2}{\mu^2}\right)+\frac{2}{s^2}(s+M^2)^2
 \left[\frac{\pi^2}{6}-{\mathrm{Li_2}}\left(1+\frac{s}{M^2}\right)\right]
\right\} \nonumber \\ & \times&  
 \left[1+\frac{2t}{s}+\lambda_1\lambda_3 \right], 
\end{eqnarray}
where
\begin{equation} {\mathrm{Li_2}}(x) \ = \ {\mathrm{dilog}}(1-x) \ =
 \ -\int_0^x \frac{\ln(1-y)}{y}dy. \end{equation}
We have assumed here that all the quark masses are negligible. In the case
where the outgoing quarks are $b$-quarks and the gauge boson in the loop
is a $W$  though, the internal quark is a $t$-quark, whose mass must be taken into
consideration. In such cases the contribution to the amplitude is not readily
expressed as an analytic function and the contribution is
calculated numerically using Veltman-Passarino reduction \cite{VP} and the FF
library \cite{FF}.
 
\bigskip \bigskip

\subsubsection{Box graphs}

The contribution to the amplitude from the box graph with two massless
gauge bosons exchanged in $s$-channel \\

\centerline{\epsfig{file=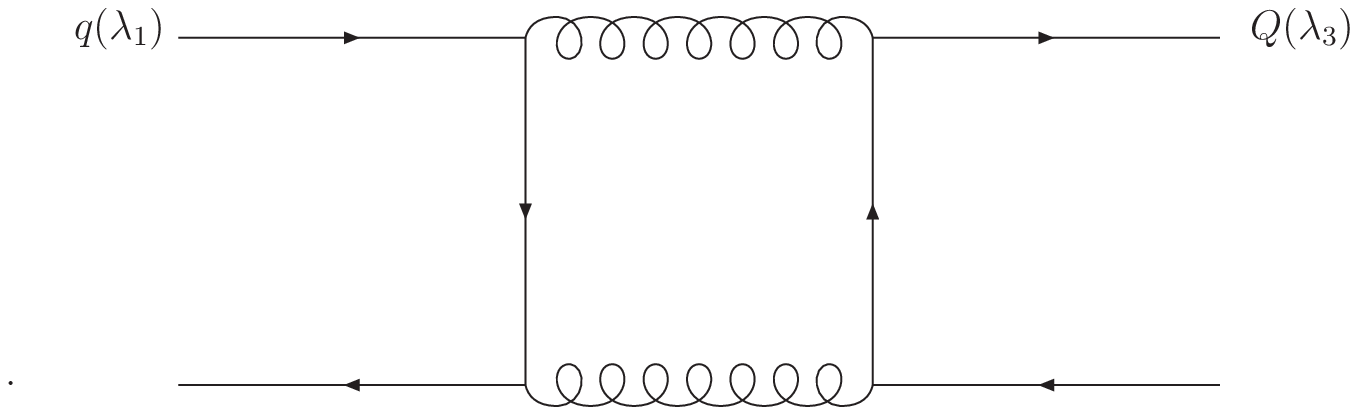, width=7 cm}}

\noindent
is
\begin{eqnarray}
\frac{1}{16\pi^2} &\times &\Bigg\{ \frac{t}{s} \left[
4\pi^2-4\ln^2\left(\frac{-s}{\mu^2}\right)
 +8\ln\left(\frac{s}{t}\right)\ln\left(\frac{-s}{\mu^2}\right)
 \right]
\nonumber \\ &+ &
(1+\lambda_1\lambda_3)\Bigg[-\frac{s}{u}\pi^2
+4\ln\left(\frac{s}{t}\right) 
\ln\left(\frac{-s}{\mu^2}\right)-\left(2+\frac{s}{u}
\right)\ln^2\left(\frac{s}{t}\right)\nonumber \\ & 
-& 2\ln\left(\frac{s}{t}\right) - 2 \ln^2\left(\frac{-s}{\mu^2}\right)
\Bigg]\Bigg\}. \end{eqnarray}
\bigskip

The contribution to the amplitude from the box graph with a massless
and a massive gauge boson exchanged in $s$-channel \\

\centerline{\epsfig{file=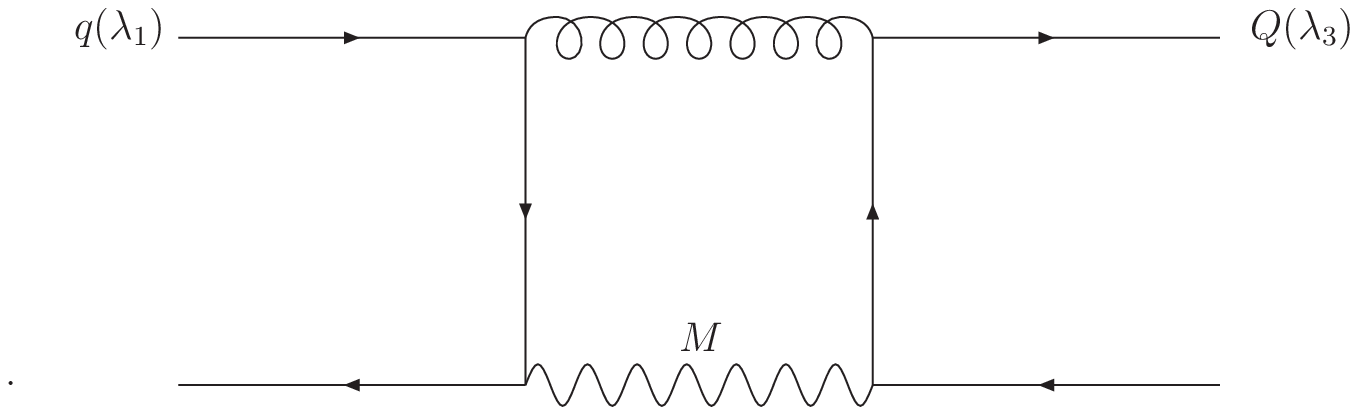, width=7 cm}}

\noindent
is
\begin{eqnarray}
\frac{1}{4\pi^2(s-M^2)}& \times& \Bigg\{ -4 \frac{tM^2}{s} 
\ln\left(1-\frac{s}{M^2} \right)
\ln\left(\frac{M^2-s}{\mu^2} \right)
-4t\ln\left(\frac{-t}{M^2}\right)
 \ln\left(1-\frac{s}{M^2}\right) \nonumber \\
 &+& 2t\ln\left(\frac{-t}{M^2}\right)\ln\left(1+\frac{t}{M^2}\right)
-t \ln^2\left( \frac{-t}{\mu^2}\right)  \nonumber \\ &-& 
 4t\left(1+\frac{s}{M^2}\right)
 {\mathrm{Li_2}}\left(\frac{s}{M^2}\right) 
 + 2 t {\mathrm{Li_2}}\left(\frac{-t}{M^2}\right) \nonumber \\ & +&
  (1+\lambda_1\lambda_3)\Bigg[ 
-2M^2 \ln\left(1-\frac{s}{M^2}\right)\ln\left(\frac{M^2-s}{\mu^2}\right)
-\frac{(s-M^2)}{s}\ln\left(1-\frac{s}{M^2}\right) \nonumber \\ &- & 
\frac{s}{2}\ln^2\left(\frac{-t}{\mu^2}\right)
 +\left(\frac{(s-M^2)^2}{u}-2M^2\right)\ln\left(\frac{-t}{M^2}\right)
 \ln\left(\frac{M^2-s}{t+M^2}\right) \nonumber \\ &- &
s\ln\left(\frac{-t}{M^2}\right)\ln\left(1+\frac{t}{M^2}\right)
+(s-M^2)\ln\left(\frac{-t}{M^2}\right) \nonumber \\ & +& 
\left(\frac{(s-M^2)^2}{u}-4M^2\right) 
\left[ {\mathrm{Li_2}}\left(\frac{s}{M^2}\right)-{\mathrm{Li_2}}\left(\frac{-t}{M^2}\right)
\right]
\nonumber \\ & -& 
\left(s+2M^2\right){\mathrm{Li_2}}\left(\frac{-t}{M^2}\right)
\Bigg]
\Bigg\}.  
\end{eqnarray}

Note that the terms containing $u$ in the denominator 
are always multiplied by a coefficient that vanishes when $s=-t$,
so that there is no bogus singularity in the backward direction.
The contributions from box graphs in which the 
 two gauge bosons are crossed in $s$-channel may be obtained
 from the above box contributions by using the substitutions
 $t \leftrightarrow u$ and $\lambda_3 \rightarrow -\lambda_3$.
In the end, 
all virtual corrections to all partonic processes with external
quarks and/or anti-quarks can be obtained from these prototype
graphs and expressions by multiplying by the appropriate couplings and colour factors
and by using the appropriate crossing relations. For reason of space we
do not report here all such results. Rather, we make available upon request
the codes that implement them.

\subsection*{Acknowledgements}
SM thanks Joey Huston, Nigel Glover, Denis Comelli and Paolo Ciafaloni for 
useful discussions. All authors acknowledge useful email exchanges and discussions
with Robert Thorne.

\clearpage

\begin{figure}[h]

\vspace*{-0.5truecm}
\caption{\small The set of interferences that contributes to process 1 ($gg\rightarrow q\bar{q}$) and, if we reverse
the direction of time, 2 ($q\bar{q}\rightarrow gg$).}
\label{fig:proc1and2a}
\end{figure}
\clearpage
\begin{figure}[h]
  %
\vspace*{-0.5truecm}
\caption{\small Continuing the set of interferences that contribute to process 1 ($gg\rightarrow q\bar{q}$) and, if we reverse
the direction of time, 2 ($q\bar{q}\rightarrow gg$).}
\label{fig:proc1and2b}
\end{figure}
\clearpage
\begin{figure}[h]
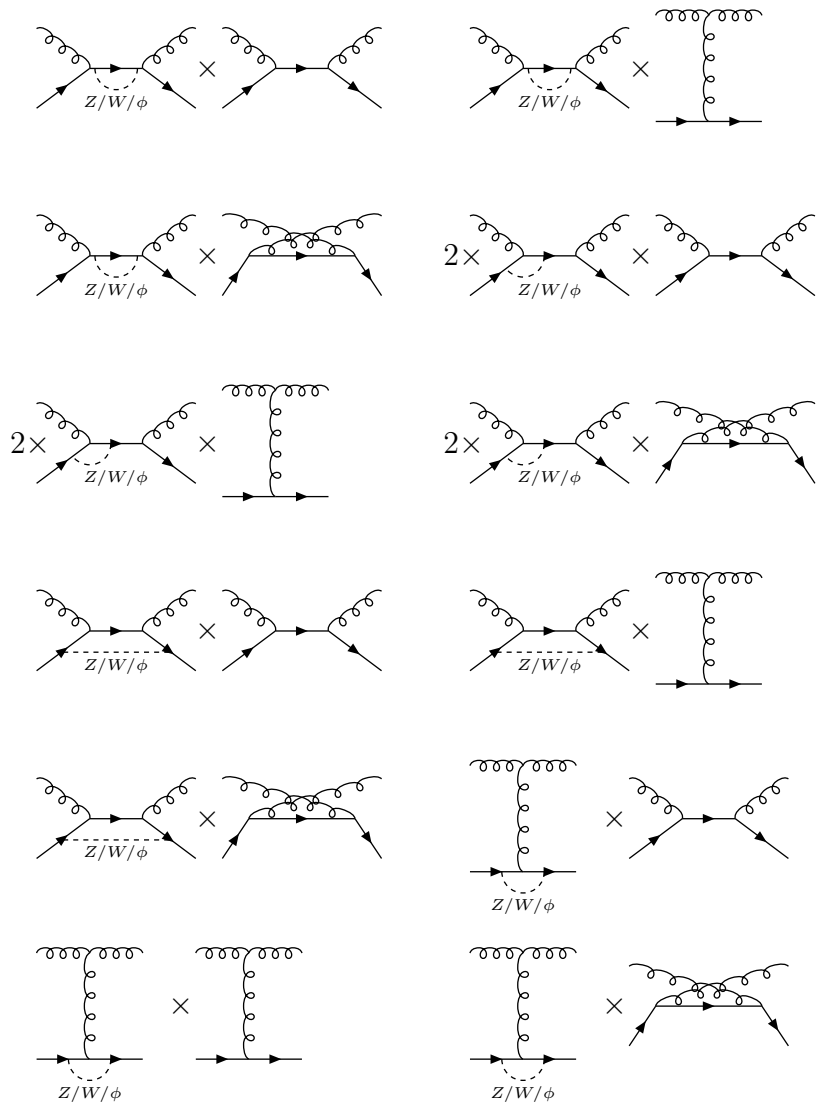

  %
\caption{\small The set of interferences that contribute to process 3 ($qg\rightarrow qg$) and, if we reverse
the direction of the fermion arrow, 4 ($\bar{q}g\rightarrow \bar{q}g$).}
\label{fig:proc3and4a}
\end{figure}
\clearpage
\begin{figure}[h]
  %
\vspace*{-0.5truecm}
\caption{\small Continuing the set of interferences that contribute to process 3 ($qg\rightarrow qg$) and, if we reverse
the direction of the fermion arrow, 4 ($\bar{q}g\rightarrow \bar{q}g$).}
\label{fig:proc3and4b}
\end{figure}
\clearpage
\begin{figure}
%
\caption{\small The set of interferences that contributes to process 5 ($qq\rightarrow qq$) and 6
($\bar{q}\bar{q}\rightarrow \bar{q}\bar{q}$), the latter obtained by reversing the arrows on all fermion lines
of the former.}
\label{fig:proc5and6a}
\end{figure}
\clearpage
\begin{figure}
%
\vspace*{-0.5truecm}
\caption{\small Continuing the set of interferences that contributes to process 5 ($qq\rightarrow qq$) and 6
($\bar{q}\bar{q}\rightarrow \bar{q}\bar{q}$), the latter obtained by reversing the arrows on all fermion lines
of the former.}
\label{fig:proc5and6b}
\end{figure}
\clearpage
\begin{figure}
%
\caption{\small The set of interferences that contributes to process 7 ($qQ\rightarrow qQ$ (same generation)) and 8
($\bar{q}\bar{Q}\rightarrow \bar{q}\bar{Q}$ (same generation)), the latter obtained by reversing the arrows on all fermion lines
of the former. Here, the
weak couplings to the two fermion lines will be different, this means we cannot implement the last 
two interferences as a factor of 2 as we could in Figs.~\ref{fig:proc1and2a}, \ref{fig:proc1and2b}.}
\label{fig:proc7and8a}
\end{figure}
\clearpage
\begin{figure}
%
\vspace*{-0.5truecm}
\caption{\small The set of interferences that contributes to  process 9 ($qQ\rightarrow qQ$ (different generation)) and 10
($\bar{q}\bar{Q}\rightarrow \bar{q}\bar{Q}$ (different generation)), the latter obtained by reversing the arrows on all fermion lines
of the former. Here, the
weak couplings to the two fermion lines will be different, this means we cannot implement the last 
two interferences as a factor of 2 as we could in Figs.~\ref{fig:proc1and2a}, \ref{fig:proc1and2b}.}
\label{fig:proc9and10a}
\end{figure}
\clearpage
\begin{figure}
  %
\vspace*{-0.5truecm}
\caption{\small The set of interferences that contributes to  process 11 ($q\bar{q}\rightarrow q\bar{q}$).}
\label{fig:proc11a}
\end{figure}
\clearpage
\begin{figure}
  %
\vspace*{-0.5truecm}
\caption{\small The set of interferences that contributes to  process 11 ($q\bar{q}\rightarrow q\bar{q}$): continued from
Fig.~\ref{fig:proc11a}.}
\label{fig:proc11b}
\end{figure}
\clearpage
\begin{figure}
  %
\vspace*{-0.5truecm}
\caption{\small The set of interferences that contributes to  process 11 ($q\bar{q}\rightarrow q\bar{q}$): continued from
Figs.~\ref{fig:proc11a} and \ref{fig:proc11b}.}
\label{fig:proc11c}
\end{figure}
\clearpage
\begin{figure}
  %
\vspace*{-0.5truecm}
\caption{\small The set of interferences  that contributes to process 12 ($q\bar{q}\rightarrow Q\bar{Q}$ (same generation)).}
\label{fig:proc12a}
\end{figure}
\clearpage
\begin{figure}
  %
\vspace*{-0.5truecm}
  \caption{\small The set of interferences  that contributes to process 13 ($q\bar{q}\rightarrow Q\bar{Q}$ (different generation)).}
  \label{fig:proc13a}
\end{figure}
\clearpage
\begin{figure}
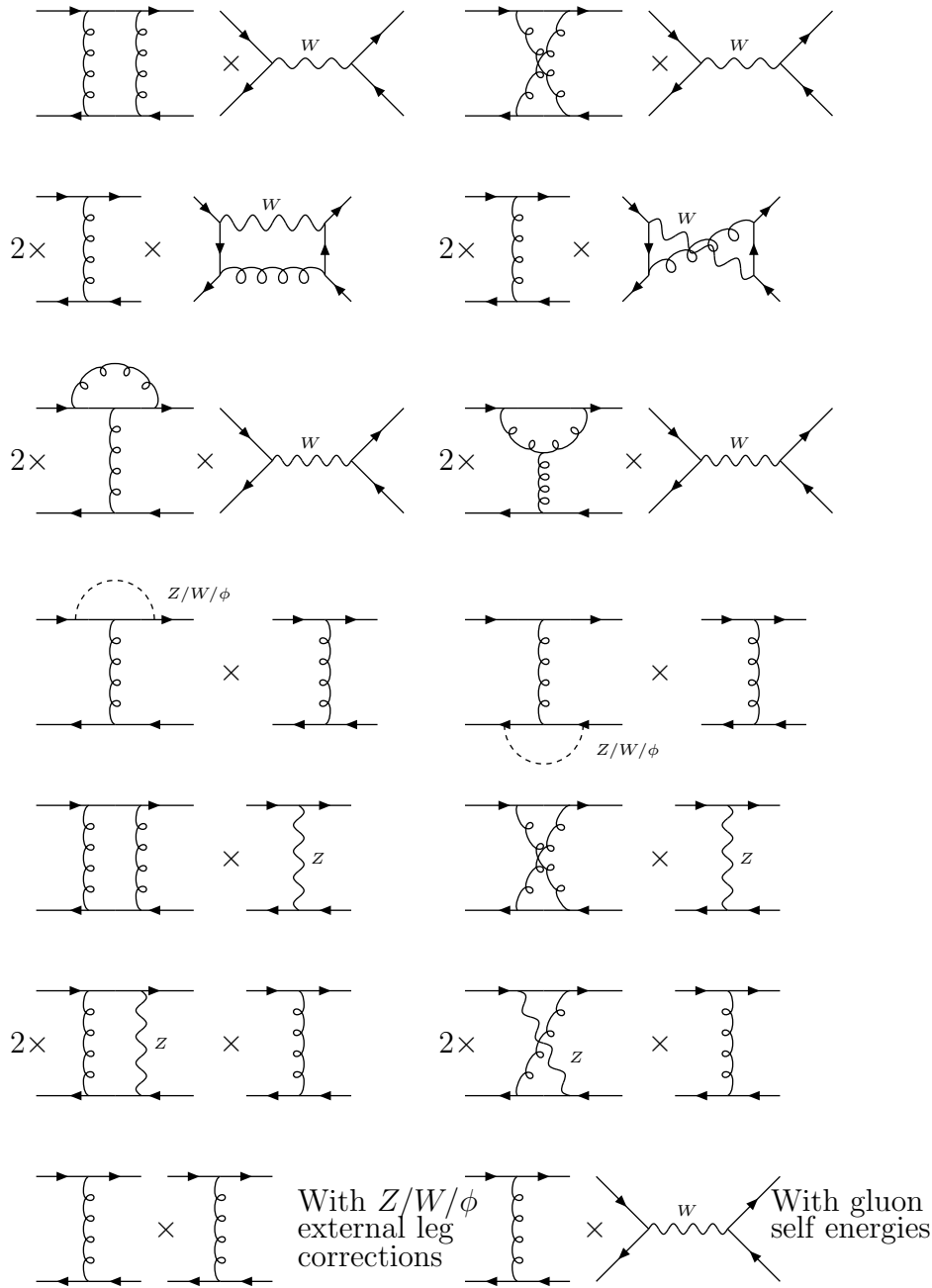

  %
\vspace*{-0.5truecm}
\caption{\small The set of interferences  that contributes to  process 14 ($q\bar{Q}\rightarrow q\bar{Q}$ (same generation)).}
\label{fig:proc14a}
\end{figure}
\clearpage
\begin{figure}
  %
\vspace*{-0.5truecm}
\caption{\small The set of interferences  that contributes to process 15 ($q\bar{Q}\rightarrow q\bar{Q}$ (different generation)).}
\label{fig:proc15a}
\end{figure}
\begin{figure}
%
\vspace*{-0.5truecm}
\caption{\small The interferences that contribute to process 16 (diagrams with a $Z$ exchange, $qg\rightarrow qq\bar{q}$) and 18 
(diagrams with a $W$ exchange, $qg\rightarrow qQ\bar{Q}$). If we reverse the direction of the incoming fermion line we also obtain process
17 ($\bar{q}g\rightarrow \bar{q}\bar{q}q$) and 19 ($\bar{q}g\rightarrow \bar{q}\bar{Q}Q$).}
\label{fig:proc16to19a}
\end{figure}

\clearpage

\begin{figure}[htb]
\label{fig_Tevatron_PDFs_newscale}
\begin{center}
\hspace*{0.005truecm}{\epsfig{file=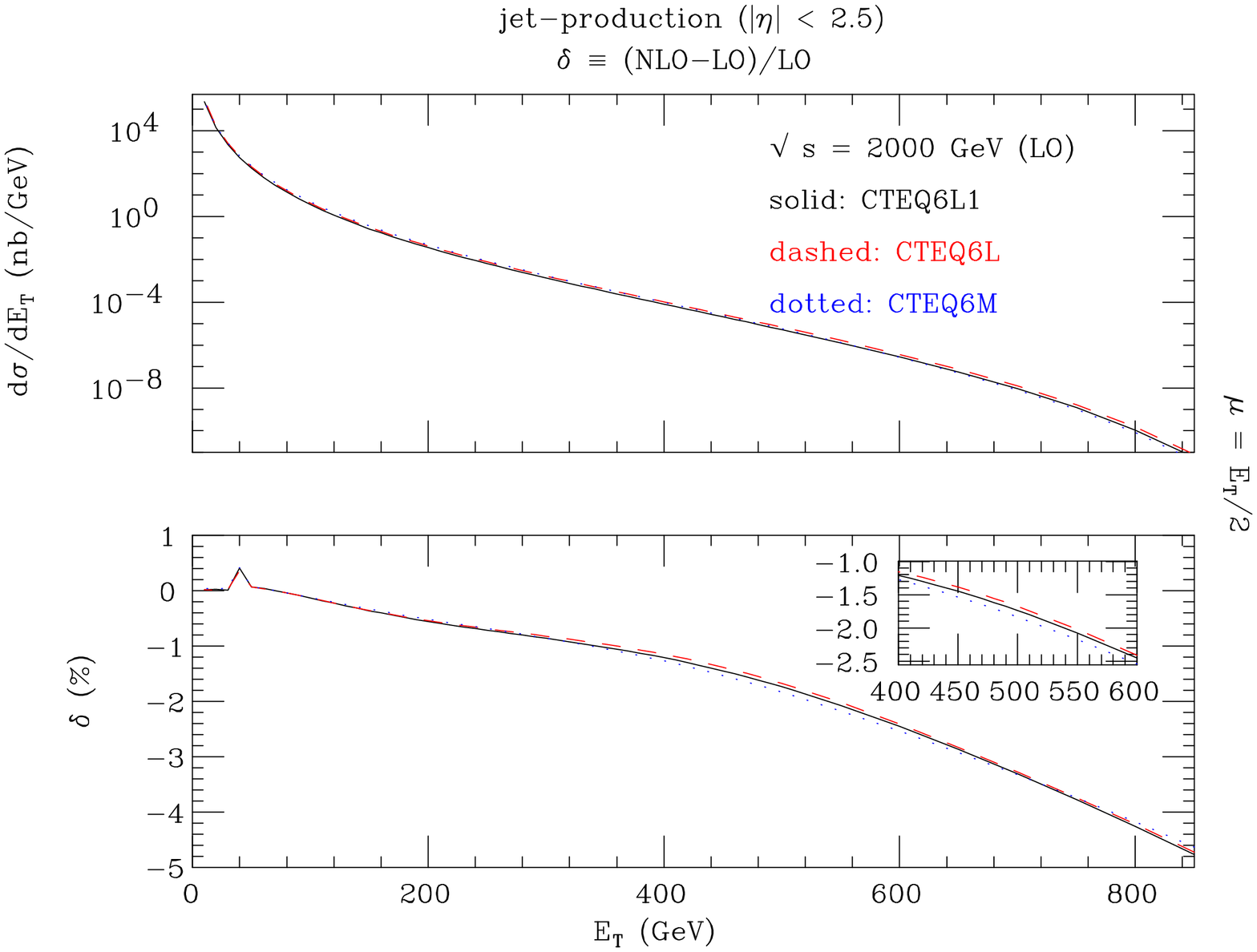,height=160mm,angle=90}}
\caption{The  effects of the 
${\cal O}(\alpha_{\mathrm{S}}^2\alpha_{\mathrm{W}})$
corrections [bottom] relative to the full LO results
(i.e., through
${\cal O}(\alpha_{\mathrm{S}}^2+\alpha_{\mathrm{S}}\alpha_{\mathrm{EW}}+\alpha_{\mathrm{EW}}^2)$) [top]
for the case of Tevatron (Run 2) for three choices of PDFs. 
They are plotted as function of the jet transverse
energy $E_T$. The cut $|\eta|<2.5$
has been enforced, alongside the standard jet cone requirement $\Delta R>0.7$.
The factorisation/renormalisation scale adopted was $\mu=\mu_F\equiv\mu_R=E_T/2$.}
\end{center}
\end{figure}

\clearpage

\begin{figure}[htb]
\label{fig_Tevatron_scales}
\begin{center}
\hspace*{0.005truecm}{\epsfig{file=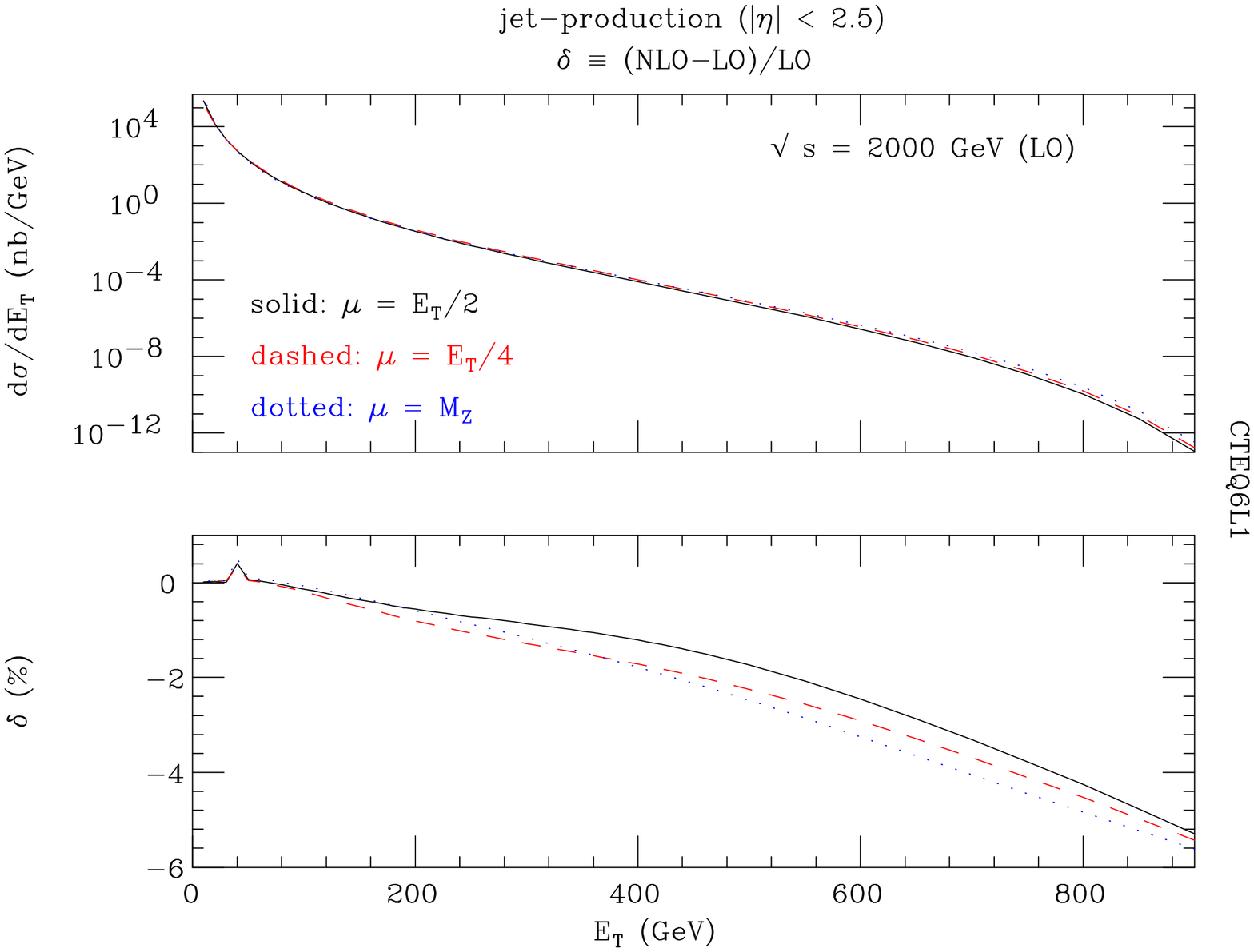,height=160mm,angle=90}}
\caption{The  effects of the 
${\cal O}(\alpha_{\mathrm{S}}^2\alpha_{\mathrm{W}})$
corrections [bottom] relative to the full LO results
(i.e., through
${\cal O}(\alpha_{\mathrm{S}}^2+\alpha_{\mathrm{S}}\alpha_{\mathrm{EW}}+\alpha_{\mathrm{EW}}^2)$) [top]
for the case of Tevatron (Run 2) for three choices of
factorisation/renormalisation scale. 
They are plotted as function of the jet transverse
energy $E_T$. The cut $|\eta|<2.5$
has been enforced, alongside the standard jet cone requirement $\Delta R>0.7$.
The PDFs used were CTEQ6L1.}
\end{center}
\end{figure}

\clearpage

\begin{figure}[htb]
\label{fig_LHC_PDFs_newscale}
\begin{center}
\hspace*{0.005truecm}{\epsfig{file=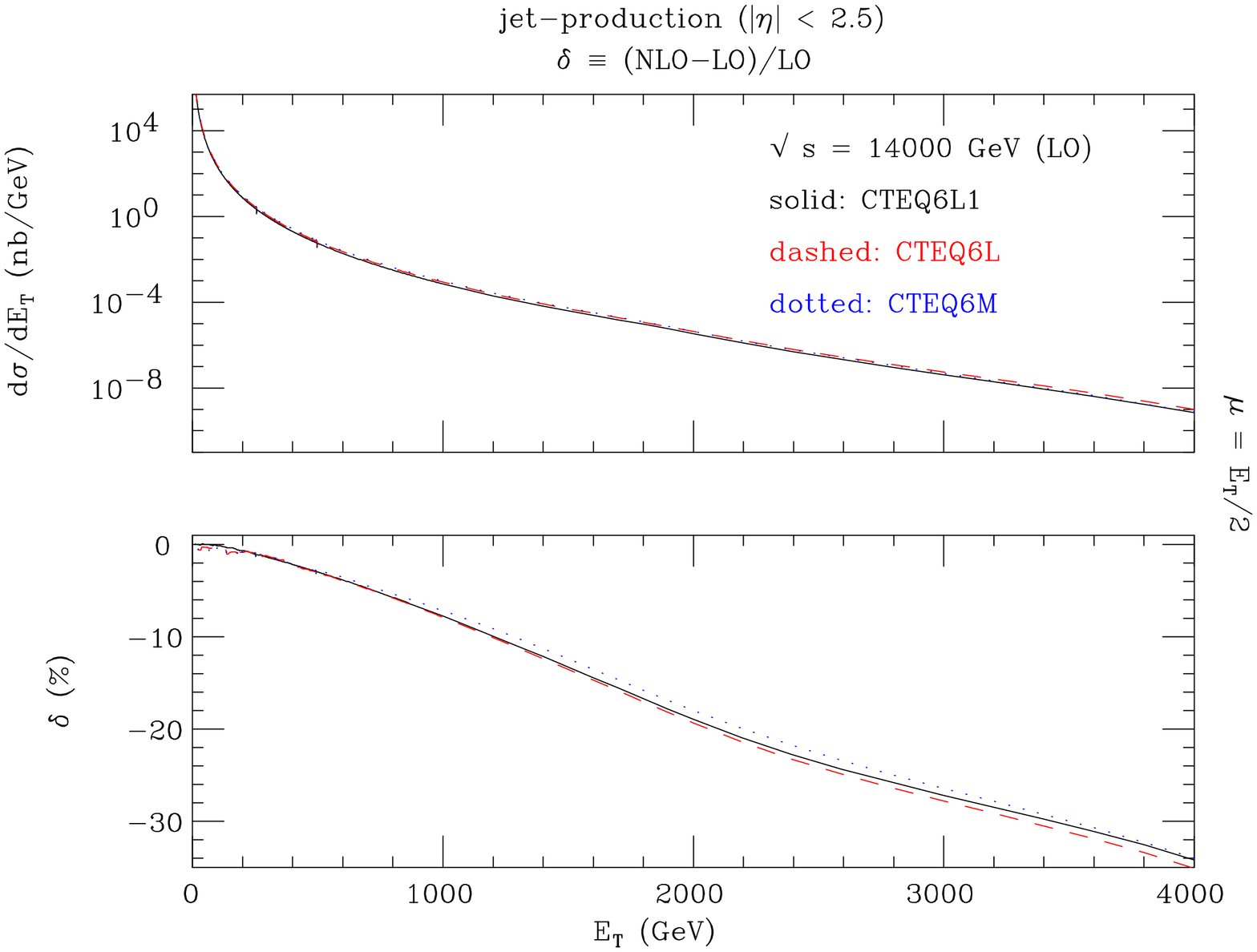,height=160mm,angle=90}}
\caption{The  effects of the 
${\cal O}(\alpha_{\mathrm{S}}^2\alpha_{\mathrm{W}})$
corrections [bottom] relative to the full LO results
(i.e., through
${\cal O}(\alpha_{\mathrm{S}}^2+\alpha_{\mathrm{S}}\alpha_{\mathrm{EW}}+\alpha_{\mathrm{EW}}^2)$) [top]
for the case of LHC for three choices of PDFs. 
They are plotted as function of the jet transverse
energy $E_T$. The cut $|\eta|<2.5$
has been enforced, alongside the standard jet cone requirement $\Delta R>0.7$.
The factorisation/renormalisation scale adopted was $\mu=\mu_F\equiv\mu_R=E_T/2$.}
\end{center}
\end{figure}

\clearpage

\begin{figure}[htb]
\label{fig_LHC_scales}
\begin{center}
\hspace*{0.005truecm}{\epsfig{file=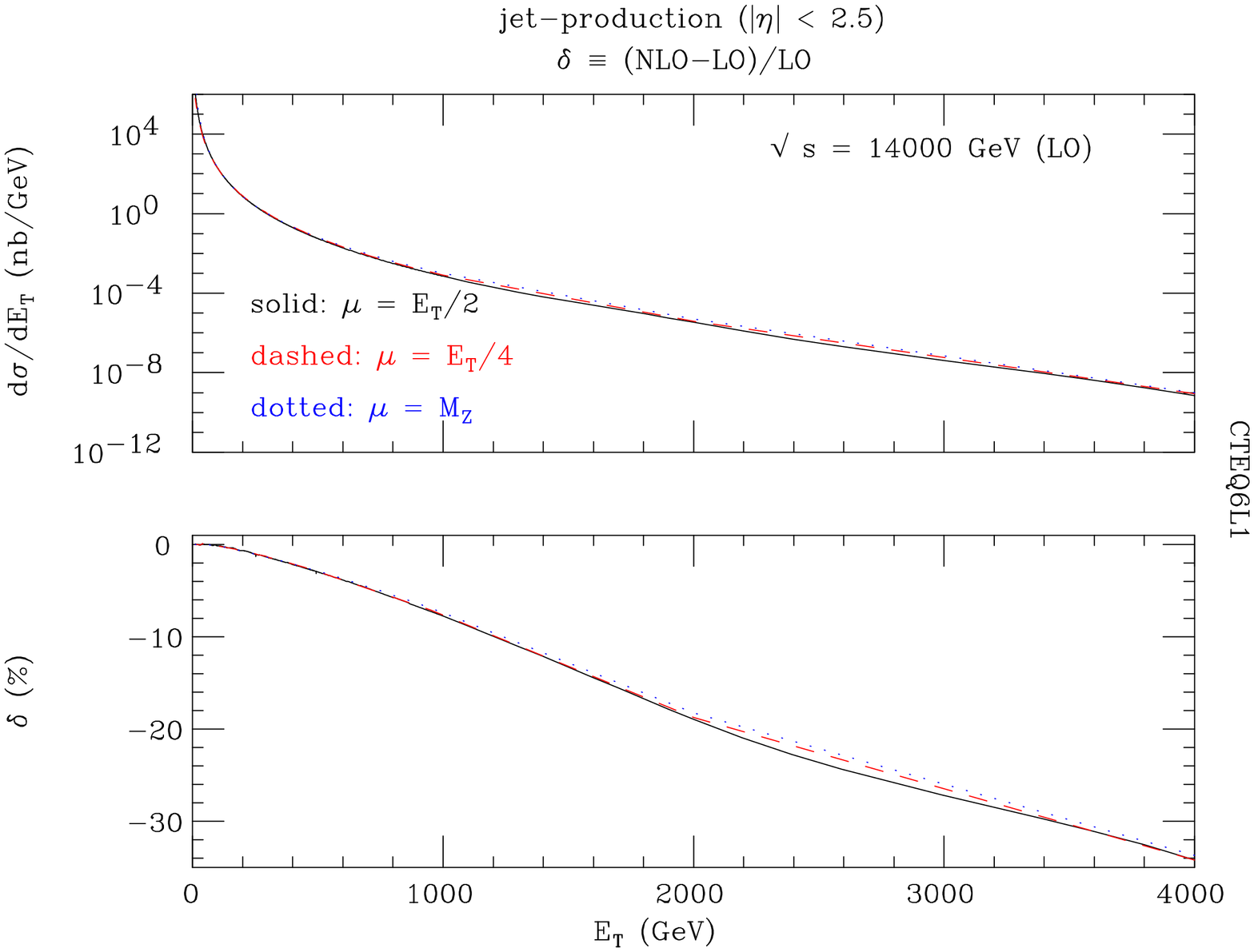,height=160mm,angle=90}}
\caption{The  effects of the 
${\cal O}(\alpha_{\mathrm{S}}^2\alpha_{\mathrm{W}})$
corrections [bottom] relative to the full LO results
(i.e., through
${\cal O}(\alpha_{\mathrm{S}}^2+\alpha_{\mathrm{S}}\alpha_{\mathrm{EW}}+\alpha_{\mathrm{EW}}^2)$) [top]
for the case of LHC for three choices of
factorisation/renormalisation scale. 
They are plotted as function of the jet transverse
energy $E_T$. The cut $|\eta|<2.5$
has been enforced, alongside the standard jet cone requirement $\Delta R>0.7$.
The PDFs used were CTEQ6L1.}
\end{center}
\end{figure}

\clearpage

\begin{figure}[htb]
\label{fig_LHC_newscale_NLO}
\begin{center}
\hspace*{0.005truecm}{\epsfig{file=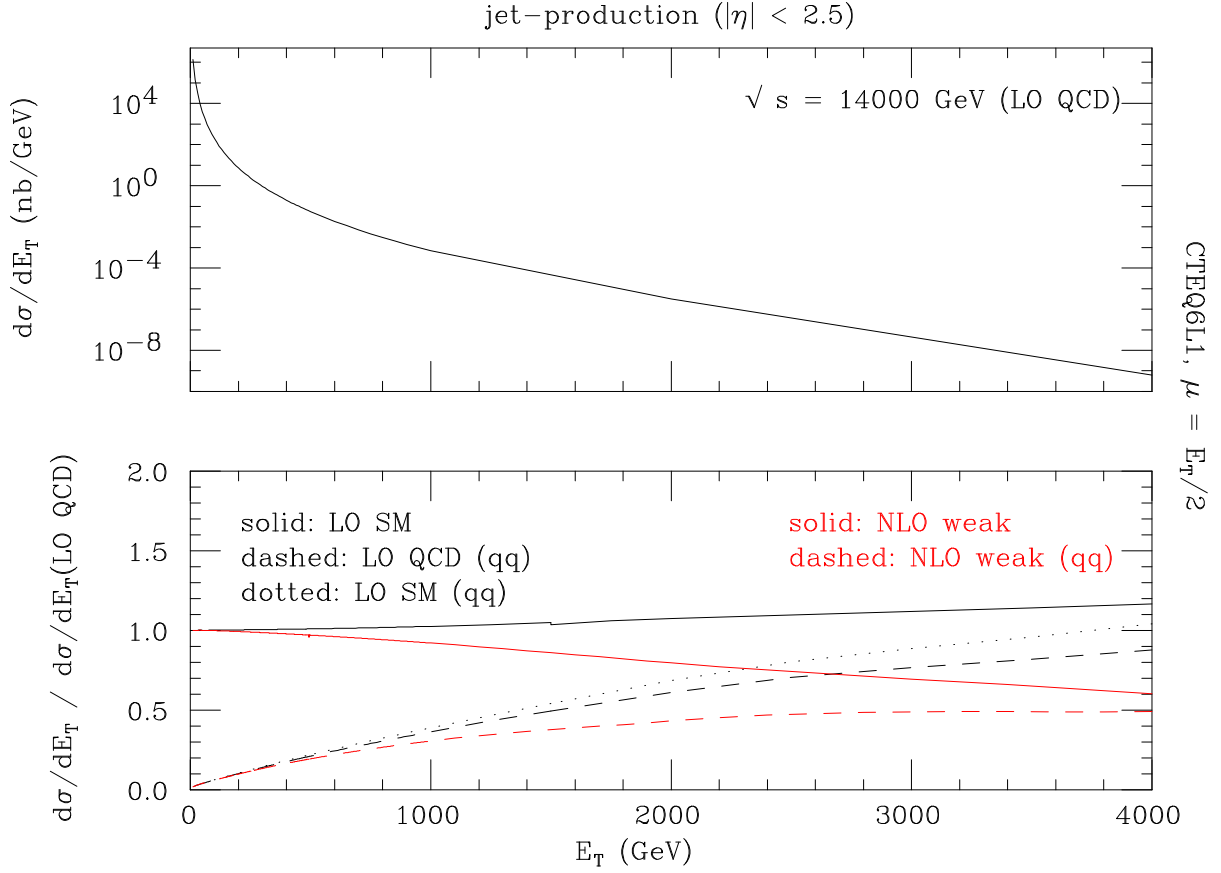,height=160mm,angle=90}}
\caption{Top: The total single jet inclusive distribution in transverse energy through
${\cal O}(\alpha_{\mathrm{S}}^2)$ at the LHC. Bottom: The  effects of the one-loop
${\cal O}(\alpha_{\mathrm{S}}^2\alpha_{\mathrm{W}})$ and 
tree-level ${\cal O}(\alpha_{\mathrm{S}}\alpha_{\mathrm{EW}}
+\alpha^2_{\mathrm{EW}})$ 
corrections relative to the spectrum above. The label {\tt (qq)} refers to the case
of subprocesses with no gluons in the initial state. Rates are plotted as function of the jet transverse
energy $E_T$. The cut $|\eta|<2.5$
has been enforced, alongside the standard jet cone requirement $\Delta R>0.7$.
The PDFs used were CTEQ6L1 whilst the factorisation/renormalisation scale adopted was $\mu=\mu_F\equiv\mu_R=E_T/2$.}
\end{center}
\end{figure}

\clearpage

\begin{figure}[!t]
\begin{center}
\vspace*{-1.25truecm}
\hspace*{0.005truecm}{\epsfig{file=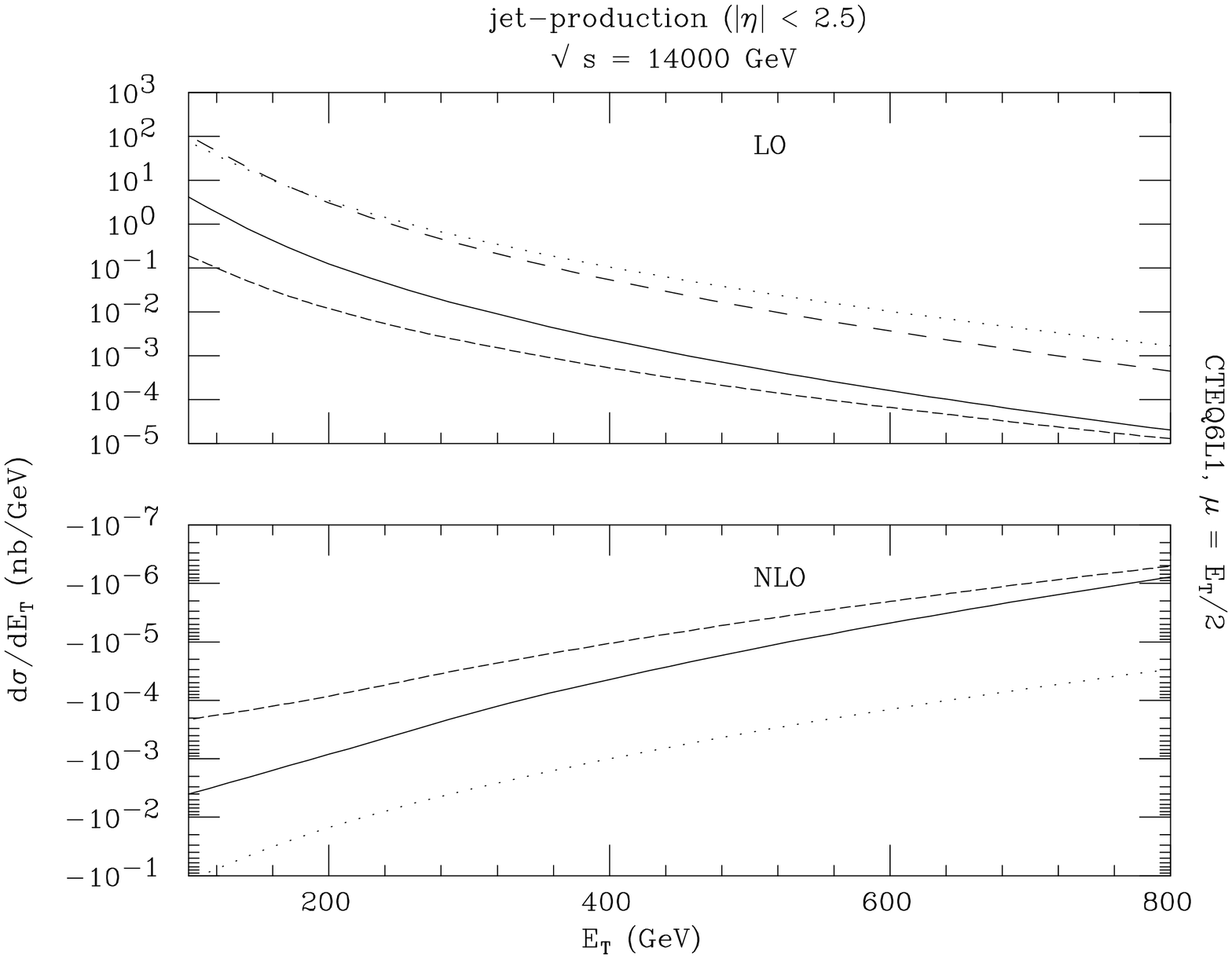,height=115mm,angle=90}}
\end{center}
\centerline{(a)}
\end{figure}
\begin{figure}[!b]
\begin{center}
\hspace*{0.005truecm}{\epsfig{file=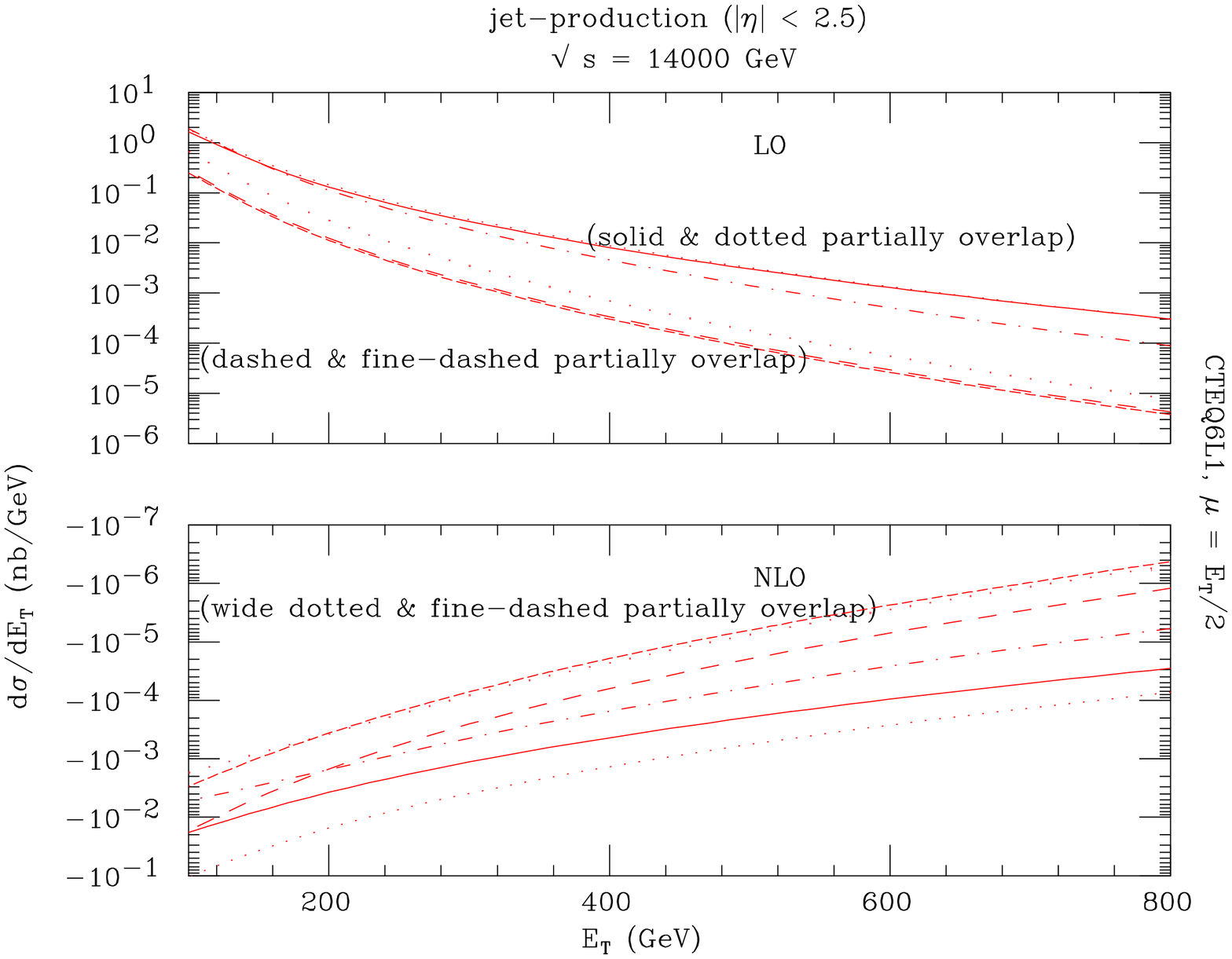,height=115mm,angle=90}}
\end{center}
\centerline{(b)}
\end{figure}

\clearpage

\begin{figure}[!t]
\label{fig_LHC_newscale_iproc}
\begin{center}
\vspace*{-1.25truecm}
\hspace*{0.005truecm}{\epsfig{file=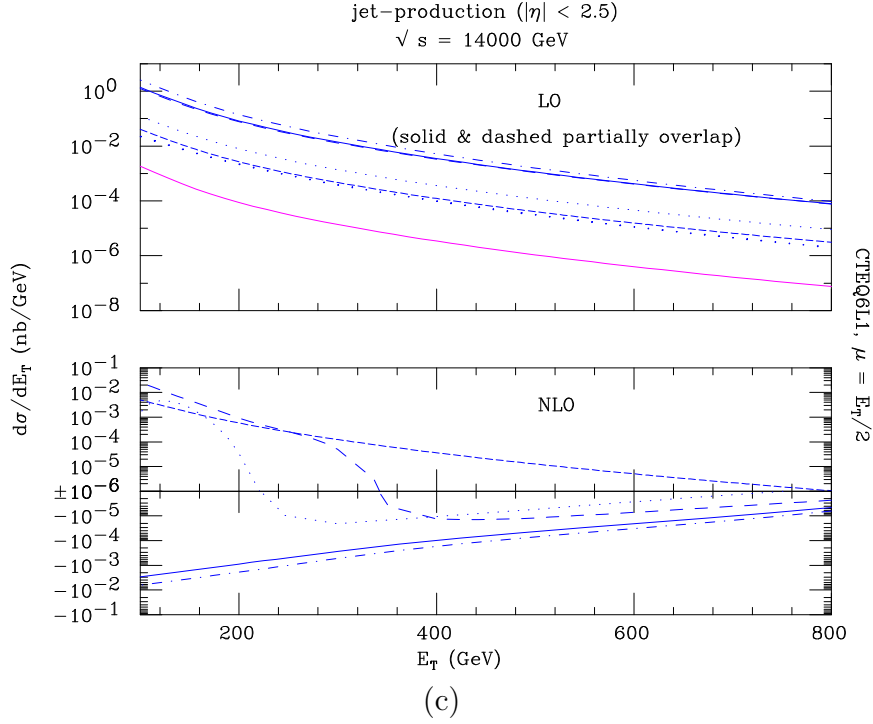,height=115mm,angle=90}}
\centerline{(c)}
\vspace*{-0.75truecm}
\caption{The total single jet inclusive distribution in transverse energy through tree-level (LO)
via ${\cal O}(\alpha_{\mathrm{S}}^2)$, ${\cal O}(\alpha_{\mathrm{S}}\alpha_{\mathrm{EW}})$
and ${\cal O}(\alpha_{\mathrm{EW}}^2)$ (top) and one-loop (NLO) corrections  at 
${\cal O}(\alpha_{\mathrm{S}}^2\alpha_{\mathrm{W}})$ 
at the LHC. 
The cut $|\eta|<2.5$
has been enforced, alongside the standard jet cone requirement $\Delta R>0.7$.
The PDFs used were CTEQ6L1 whilst the factorisation/renormalisation scale adopted was $\mu=\mu_F\equiv\mu_R=E_T/2$.
Absolute rates are given for each of the suprocesses (1)--(15), labelled as follows (round brakets imply
summation over corresponding channels):}
\end{center}
\end{figure}
\vspace*{-1.95truecm}
\noindent
\begin{eqnarray}\nonumber
{\rm{(a)}}~~~~~~~~~~g g &\to& q \bar q~({\rm{solid~black,~LO~\&~NLO}}) \\ \nonumber 
{\rm{(a)}}~~~~~~~~~~q \bar q &\to& g g~({\rm{fine-dashed~black,~LO~\&~NLO}}) \\ \nonumber 
{\rm{(a)}}~~~~~~q(\bar q) g &\to& q(\bar q) g~({\rm{dotted~black,~LO~\&~NLO}}) \\ \nonumber 
{\rm{(a)}}~~~~~~~~~~gg &\to& gg~({\rm{dashed~black,~LO~only}}) \\ \nonumber 
{\rm{(b)}}~~~~~~~~~~q q &\to& q q~({\rm{solid~red,~LO~\&~NLO}}) \\ \nonumber 
{\rm{(b)}}~~~~~~~~~~\bar q \bar q &\to& \bar q \bar q~({\rm{fine-dashed~red,~LO~\&~NLO}}) \\ \nonumber 
{\rm{(b)}}~~~~~~~~~q Q &\to& q Q~({\rm{same~generation}})~({\rm{dotted~red,~LO~\&~NLO}}) \\ \nonumber 
{\rm{(b)}}~~~~~~~~~\bar q \bar Q &\to& \bar q \bar Q~({\rm{same~generation}})~({\rm{dashed~red,~LO~\&~NLO}}) \\ \nonumber 
{\rm{(b)}}~~~~~~~~~q Q &\to& q Q~({\rm{different~generation}})~({\rm{dot-dashed~red,~LO~\&~NLO}}) \\ \nonumber 
{\rm{(b)}}~~~~~~~~~\bar q \bar Q &\to& \bar q \bar Q~({\rm{different~generation}})~({\rm{wide-dotted~red,~LO~\&~NLO}}) \\ \nonumber 
{\rm{(c)}}~~~~~~~~~~q \bar q &\to& q \bar q~({\rm{solid~blue,~LO~\&~NLO}}) \\ \nonumber 
{\rm{(c)}}~~~~~~~~~~q \bar q &\to& Q \bar Q~({\rm{same~generation}})~({\rm{fine-dashed~blue,~LO~\&~NLO}}) \\ \nonumber 
{\rm{(c)}}~~~~~~~~~~q \bar q &\to& Q \bar Q~({\rm{different~generation}})~({\rm{dotted~blue,~LO~\&~NLO}}) \\ \nonumber 
{\rm{(c)}}~~~~~~~~~q \bar Q &\to& q \bar Q~({\rm{same~generation}})~({\rm{dashed~blue,~LO~\&~NLO}}) \\ \nonumber 
{\rm{(c)}}~~~~~~~~~q \bar Q &\to& q \bar Q~({\rm{different~generation}})~({\rm{dot-dashed~blue,~LO~\&~NLO}}) \\ \nonumber 
{\rm{(c)}}~q(\bar q) q'(\bar q') &\to& Q(\bar Q) Q'(\bar Q')~({\rm{wide-dotted~blue,~LO~only}}) \\ \nonumber 
{\rm{(c)}}~~~~~~~~~q \bar q'&\to& Q \bar Q'~({\rm{solid~purple,~LO~only}}) \nonumber 
\end{eqnarray}

\clearpage

\begin{table}[h]
    \begin{center}
$\sqrt s=14$  TeV, $E_T=800$ GeV \\
    \begin{tabular}{|l||c|c|c|} \hline
     Subprocess & (a) & (b)& (c)  \\ \hline    
        $gg\to gg$     & & & 14.3\\ \hline 
               (1)     &--0.0292 &--3.88 &0.643\\ \hline 
               (2)     &--0.0189 & --3.97 & 0.408\\ \hline 
          (3)--(4)     &--1.089 & --1.75 & 53.2\\ \hline 
          (5)--(6)     &--1.078 & --9.36 & 9.85\\ \hline 
          (7)--(8)     &--2.78 & --23.9 & 9.93\\ \hline 
         (9)--(10)     &--0.239 & --6.68& 3.07\\ \hline 
              (11)     &--0.169 & --5.82 & 2.48\\ \hline 
              (12)     &0.0390 & 33.9 & 0.0985\\ \hline 
              (13)     &--0.0274 & --8.09 & 0.290\\ \hline 
              (14)     &--0.0861 & --2.98 & 2.47\\ \hline 
              (15)     &--0.239 & --6.69 & 3.05\\ \hline 
$qq'\to QQ'$ or $\bar q\bar q'\to \bar Q\bar Q'$  & & & 0.0616\\ \hline  
$q\bar q'\to Q\bar Q'$                       & & & 0.00239\\ \hline 
          Total        &--5.71 & & \\ \hline 
\end{tabular}
    \caption{\small The contributions of subprocesses (1)--(15) to the total correction
through ${\cal O}(\alpha_{\mathrm{S}}^2\alpha_{\mathrm{W}})$ with respect to the
full LO result, i.e.,  through
${\cal O}(\alpha_{\mathrm{S}}^2+\alpha_{\mathrm{S}}\alpha_{\mathrm{EW}}+\alpha_{\mathrm{EW}}^2)$,
in the case of the differential cross section at LHC for $E_T=800$ GeV.
Column (a) indicates the percentage contribution of the correction to the total; 
column (b) indicates the percentage correction to the individual partonic process; 
column (c) indicates the percentage contribution from that partonic process at tree-level to the differential cross section.
Here, we have paired
together the channels with identical Feynman diagram topology.
The cut $|\eta|<2.5$
has been enforced, alongside the standard jet cone requirement $\Delta R>0.7$.
The PDFs used were CTEQ6L1 whilst the factorisation/renormalisation scale adopted was $\mu=\mu_F\equiv\mu_R=E_T/2$.
}
    \end{center}
\label{tab:LHC}
\end{table}

\end{document}